\documentclass[pra,showpacs,onecolumn,floatfix, groupaddress,superscriptaddress,footinbib]{revtex4-1}
\usepackage{times,amsmath,amsfonts,amssymb,latexsym}
\usepackage{graphicx,epsf}
\usepackage{subfigure}
\newcommand{\braket}[2]{\left \langle #1 | #2 \right\rangle}

\newcommand{\be}{\begin{equation}}
\newcommand{\ee}{\end{equation}}
\newcommand{\ba}{\begin{eqnarray}}
\newcommand{\ea}{\end{eqnarray}}
\newcommand\tr{{\mbox{Tr\,}}}
\newcommand{\ignore}[1]{}

\newcommand{\ket}[1]{\left | {#1} \right \rangle }

\newcommand{\bra}[1]{\left \langle {#1} \right | }

\def\CC{{\rm\kern.24em \vrule width.04em height1.46ex depth-.07ex
    \kern-.30em C}}
\def\P{{\rm I\kern-.25em P}}
\def\RR{{\rm
         \vrule width.04em height1.58ex depth-.0ex
         \kern-.04em R}}

\def\bbbc{{\mathchoice {\setbox0=\hbox{$\displaystyle\rm C$}\hbox{\hbox
to0pt{\kern0.4\wd0\vrule height0.9\ht0\hss}\box0}}
{\setbox0=\hbox{$\textstyle\rm C$}\hbox{\hbox
to0pt{\kern0.4\wd0\vrule height0.9\ht0\hss}\box0}}
{\setbox0=\hbox{$\scriptstyle\rm C$}\hbox{\hbox
to0pt{\kern0.4\wd0\vrule height0.9\ht0\hss}\box0}}
{\setbox0=\hbox{$\scriptscriptstyle\rm C$}\hbox{\hbox
to0pt{\kern0.4\wd0\vrule height0.9\ht0\hss}\box0}}}}
\def\bbbz{{\mathchoice {\hbox{$\sf\textstyle Z\kern-0.4em Z$}}
{\hbox{$\sf\textstyle Z\kern-0.4em Z$}}
{\hbox{$\sf\scriptstyle Z\kern-0.3em Z$}}
{\hbox{$\sf\scriptscriptstyle Z\kern-0.2em Z$}}}}

\usepackage{hyperref}

\begin{document}

\title{Ensembles of physical states and random quantum circuits on graphs}

\author{Alioscia Hamma}
\affiliation{Center for Quantum Information,
Institute for Interdisciplinary Information Sciences,
Tsinghua University, Beijing 100084, P.R. China}
\affiliation{Perimeter
Institute for Theoretical Physics, 31 Caroline St. N, N2L 2Y5,
Waterloo ON, Canada}

\author{Siddhartha Santra}
\affiliation{Department of Physics and Astronomy \& Center for Quantum Information Science and Technology,University of Southern California, Los Angeles, California 90089-0484, USA}

\author{Paolo Zanardi}
\affiliation{Department of Physics and Astronomy \& Center for Quantum Information Science and Technology,University of Southern California, Los Angeles, California 90089-0484, USA}
\begin{abstract}
In this paper we continue and extend the investigations  of the ensembles of random physical states introduced in A. Hamma \emph{et al.} [\href{http://prl.aps.org/abstract/PRL/v109/i4/e040502}{Phys. Rev. Lett. 109, 040502 (2012)}].
These ensembles are constructed by finite-length random quantum circuits (RQC) acting on the (hyper)edges of an underlying (hyper)graph structure.
The latter encodes for  the locality structure associated with finite-time quantum evolutions generated by physical i.e., local, Hamiltonians. Our goal
is to analyze physical properties of typical states in these ensembles, in particular  here we focus on proxies of quantum entanglement as purity
and $\alpha$-Renyi entropies. The problem is formulated in terms of matrix elements of superoperators which depend on the graph structure, choice
of probability measure over the  local unitaries and circuit length. In the $\alpha=2$ case these superoperators act on  a restricted   multi-qubit
space generated by permutation operators associated to the subsets of vertices of the graph. For permutationally invariant interactions the dynamics can be further restricted to an
exponentially smaller subspace. We consider different families of RQCs and study their typical entanglement properties for  finite-time as well as their asymptotic behavior. We find that area law holds in average and that the volume law is a typical property (that is, it holds in average and the fluctuations around the average are vanishing for the large system) of physical states. The area law arises when the evolution time is $O(1)$ with respect to the size $L$ of the system, while the volume law arises as typical  when the evolution time scales like $O(L)$.
\end{abstract}

\maketitle

\section{INTRODUCTION}

The study of the statistical properties of ensembles of pure quantum states is an important topic in quantum information theory, quantum statistical mechanics, and quantum many-body theory. The ensemble of pure quantum states can be chosen to mimic the uniform distribution of states in the Hilbert space, of low energy states of random Hamiltonians, or states that can be obtained by some random quantum evolution. 
One can combine group theoretical and statistical tools by construction the ensemble with group theoretic-methods. An important example in quantum information theory is the use of ensembles of random unitary operators to perform quantum algorithms. In this case one picks the unitaries from the Haar measure on the
unitary group. A related ensemble is the ensemble of states in the Hilbert space that can be obtained by some random preparation. If one is allowed to obtain all the possible states with the same probability, one has again used the Haar measure over the ensemble of states. Recently, this kind of ensembles has been studied in relation to questions of typicality of the expectation value of observables and the foundations of statistical mechanics \cite{ref1}-\cite{ref14}

In this paper, we engineer and analyze ensembles $\mathcal E$ of pure quantum states for multi-partite systems that incorporate two key features in quantum information theory: randomness and locality. In practice, $\mathcal E$ is generated by applying a Random Quantum Circuit (RQC) to a reference state. The RQC is obtained by picking stochastically subspaces of the total Hilbert space that obey some locality constraint given by an underlying graph-theoretic structure. These subspaces serve then as the support of random unitaries. In this model, randomness enters two times, in the choice of the support of the unitary and in the choice of the unitary itself. The ensemble so generated finds physical motivation in approximating the evolution of a multi-partite system given by a (time dependent) local Hamiltonian \cite{ref15, ref16}. Our analysis is conducted by putting together graph theoretic, group theoretic and operator algebra tools. In particular, we show how to encode all the relevant information about the RQC in the action of a single superoperator. 

From many perspectives, an extremely important property of pure states of multi-partite quantum systems is their entanglement \cite{ref17,ref18} (in a given bipartition $A\cup B$ of the system of linear sizes $L_A$ and $L_B$). To measure entanglement in the ensemble $\mathcal E$ we study the $\alpha-$Renyi entropies of the reduced density matrix to the subsystem $A$. In particular, because the linear entropy is a lower bound for the Von Neumann entropy, the case $\alpha=2$ is very interesting. We show in the following that the average purity in the RQC defined ensembles attains asymptotically the minimum possible value, and therefore gives a tight bound to the Von Neumann entropy.

As we shall see in the following, we are interested in studying two different regimes. The first regime is obtained when the RQC is applied in one shot, or in a number $O(1)$ of instances. The second regime, is obtained by applying the RQC a number of time scaling with the size $L_A$ of the subsystem. 

We analyze two models: (i)  the nearest neighbour {\em Random Edge Model} (REM) picks the support of the circuit in subsystems corresponding to the edges of a graph. This model leads to estimates about the time evolution of Renyi entropy  when the system evolves with Ultra-local Hamiltonians. A variant of the REM, which we call the Fully Connected REM, where the local structure is given by a completely connected graph, elucidates the utility of the superoperator approach for purity dynamics by mapping the calculations to an exponentially smaller space.  (ii) the second model of interest is the Contiguous Edge Model (CEM). In this model, we again specify a graph and then we consider a random quantum circuit which has support on the all $N$ qubits of the system (and therefore the depth of the circuit is $N$). The random circuit is thought factorizable into $k$ terms that only have support on the edges of the graph. The application of such random circuit to a reference state will here be referred to as a {\em cycle}. Cycles  can be iterated a number $n_c$ of times. The CEM is intended to produce ensembles of states that come from the unitary evolution induced by a local Hamiltonian starting from completely disentangled fiducial states $\phi$. In this case, $n_c$ represents discretized time. For regular graphs of linear size $L$, we look at the reduced state $\rho_A$ and we show in the CEM typicality of Area law of Entanglement for small times, $n_c= O(1)$ and Volume law for times $n_c= O(L_A)$. The calculations of the Purity lead to an instructive algebra of the swap operators on different subsets $A$ of the total nodes in the graph $\Gamma=(V,E)$. Moreover, asymptotically in $n_c$ we show that the reduced system becomes the completely mixed state.

Note that in the cases where the RQC is applied a number $O(L_A)$ of times, our result shows that the ensemble attains in average a  reduced system which is close to the completely mixed state. This property is shared with the ensemble of states over the Haar measure. In this sense, we show that the ensemble $\mathcal E$, even if contains a small fraction of the states in the Hilbert space, nevertheless, locally realizes the averages over the Haar measure. This concept is familiar in the context of $t-$designs. A $t-$design is an ensemble $\{ p_i, \psi_i\}$ of states  that reproduces moments over the Haar measure, that is, $\sum_i p_i (\ket{\psi}\bra{\psi})^{\otimes t}= \int_\psi (\ket{\psi}\bra{\psi})^{\otimes t}d\psi$. A unitary $t-$ design is similarly an ensemble of unitaries such that $\sum_i p_i U_i^{\otimes t}\rho (U_i^\dagger)^{\otimes t}= \int_U U^{\otimes t}\rho (U^\dagger)^{\otimes t} dU$. In other terms, the average with the Haar measure of any polynomial function of degree $t$ can be obtained by a $t-$design.

It is known that exact $t-$designs require an exponential number of states (or unitaries). On the other hand, approximate $t-$designs are much more easily obtained. 
Indeed, it has been shown in the literature \cite{ref19}-\cite{ref32} that RQC of size $n^2$ (where $n$ is the number of qubits in the system) are approximate $t-$designs, for $t=1,2,3$. The typical scheme for such RQC is to consider a random circuit where two qubit-unitaries are drawn with the Haar measure on $\mathcal U (4)$ for every pair of qubit in the system. Recently, it has been shown that even with some locality constraint RQC of linear size in $n$ are up to $3-$designs \cite{ref33}. 

It is important to emphasize that in this work, we focus on a particular problem, which allows us to obtain results beyond the asymptotic case. We are not trying to approximate {\em any} polynomial that is function of $\psi$ over the Haar measure, but, given a bipartition $A\cup B$ of the system, only the reduced system $\rho_A$ of linear size $L_A$. With this restriction, we find very powerful tools. In particular, (i) we are able to make statements about typicality of entanglement for circuits of any depth. Even the asymptotic case, scales with $L_A$ being the size of the subsystem. (ii) Our results are valid for every momentum $t$ of the statistical distribution. While the REM is a RQC with locality constraints, the protocol of the CEM is quite different, as we will see. It is motivated and inspired by the evolution induced by a distribution of time-dependent Hamiltonians. For this reason, the ensembles produced by the CEM considered here are of physical relevance for applications in the foundations of statistical mechanics.

On the other hand, when the RQC is applied a $O(1)$ number of times, we obtain an ensemble of states with typical area law for the entanglement. In some sense, this ensemble shares a lot with the set of the ground states of local Hamiltonians (without topological order). Indeed, all the ground states of such Hamiltonians can be obtained by a quantum circuit of fixed depth from some completely factorized state \cite{ref34}. Whether or not in two spatial dimensions there is an area law for the entanglement in gapped systems is an important open problem in quantum many body physics \cite{ref35}. Our approach shows, that in ensembles that contain such ground states, the area law is {\em typical}. It is known that such states have the area law as upper bound. Indeed,  the technique of the Lieb-Robinson bounds has shown that entanglement that can be produced in a subsystem $A$ by evolving for a time $t$ with a local Hamiltonian is upper bounded by $O(|\partial A|t)$ \cite{ref36},\cite{ref37}. Our study shows that such upper bounds are saturated in average, and that the fluctuations are small.

The structure of the paper is as follows: In section (\romannumeral 2) we describe the setup for our models, In section (\romannumeral 3) we briefly review essential mathematics for the remainder of the paper, In section(\romannumeral 4) and (\romannumeral 5) we present detailed studies of two particular models and conclude in section (\romannumeral 6). 

\section{SETUP}
Our scheme to investigate typicality of entanglement involves two elements:
\begin{enumerate}
\item A (hyper)graph $\Gamma=(V,E)$ whose nodes $V$ represent local Hilbert spaces corresponding to local degrees of freedom of the multi-partite system and where the edges $E$ represent the support of interactions. 
\item A Random Quantum Circuit (RQC) that acts on an input fiducial state $\phi$ that can be represented conveniently using the abstract tensor product Hilbert Space on the nodes of the Graph.
\end{enumerate}

The  system $\mathcal S$ is defined by the tensor product of many local Hilbert spaces $\mathcal H =\otimes_{v\in V} \mathcal H_v$. We are concerned with finite $d-$dimensional Hilbert spaces $\mathcal H_v$, or qudits. The total number of qudits in the sistem is $N=|V|$. We can regard the set $V$ as the set of vertices of a (hyper)graph $\Gamma = (V,E)$. We remind that a hypergraph is a set of vertices with a collection $E$ of subsets of $V$ called {\em edges}. Mathematically, $E$ is a subset of the power set $\mathcal P(V)\backslash\emptyset$. For instance, the usual graph, is a set with a collection of pairs. The hyper graph is a natural structure for multi-partite quantum  systems because we can associate the vertices to the Hilbert spaces of each particle and the edges to the support of interaction terms in the (for instance) Hamiltonian. A bipartition in $\mathcal S$ is introduced by considering a bipartition of the set of vertices $V=A\cup B $ and then considering the tensor product $\mathcal H = \mathcal H_A\otimes \mathcal H_B$, where $\mathcal H_Y = \otimes_{x\in Y\subset V} \mathcal H_x$. If $d= \dim\mathcal H_x$ for every $x$, then $\dim \mathcal H_Y = d^{|Y|}$. If $X\in E$ is an edge of the (hyper)graph, $X =\{x_1,....,x_{|X|}\}$ with $x_i\in V$ and $|X|$ is the cardinality of the edge $X$. We define the Hilbert space with support on $X$ as $\mathcal H_X=\mathcal H_{x_1}\otimes\ldots\otimes\mathcal H_{x_{|X|}}$.

One can regard the subsystem $A$ as the physical "system" of interest, and its complement $B$ as its "environment", usually assuming that $d_A\ll d_B$. 
A totally factorized state for the whole $\mathcal S$ (system +environment) can thus be written as $\ket{\phi}=\otimes_{i\in A}\ket{\phi_i}\otimes_{j\in B}\ket{\phi_j}$ 
where obviously $\ket{\phi_i}\in \mathcal H_i$.  In this work,  the fiducial state $\phi$ is any totally factorized state (and which exactly does not matter, as we shall see).

The fiducial state $\phi$ is the input to the RQC that  picks edges $X$ in $E$ to act on, according to some probability distribution $\mathcal{P}(X)$, with a unitary operator $U$ acting on $\mathcal H_X$. 
The unitary $U_X$ is picked with some measure $d\mu(U|X)$, e.g. the Haar measure over $\mathcal{U}(d^{|X|})$. In other words, we {\em first} pick an edge $X\in E$ with probability $\mathcal{P}(X)$ and then we pick a random (with measure $d\mu$) unitary with support on $\mathcal H_X$. 
The RQC itself can thus be labelled by the probability distribution of the edges and choice of the measure over unitaries. The ensemble $\mathcal E$ is then completely specified by varying on the fiducial states $\phi$ and the chosen RQC, in the following way
\be
\mathcal{E} (\mathcal{P}, d\mu) = \{ U_X\ket{\phi} \}_{X,U_X,\phi}
\ee

More generally, one can describe a general Random Quantum Circuit where instead of picking just a single edge at each step the circuit chooses a subset of the nodes. Such  RQC can be described using the joint probability distribution $\mathcal{P}^{(k)}:(X_k,X_{k-1},...,X_1)\to \mathcal{P}^{(k)}(X_k,X_{k-1},...,X_1)\in[0,1]$ where $X_i\subset V,i=1,2,...,k$ are subsets of the set of vertices $V$ of the (hyper)graph, or, in other words, edges in $E$. The sequence of \emph{set valued} Random Variables, $X_1,X_2,....,X_k$, can be seen as a stochastic process of length $`k'$. Assuming that such a selection of subsets of V is a Markovian process, one can express the action of RQCs of arbitrary depth using just the Markov Transition matrix, $M^{(k)}(X_k|X_{k-1})$ which satisfy $\sum_{X_\alpha}M^{(i)}(X_\alpha|X_\beta)=1 \forall i=1,2,...,k$ where $\alpha,\beta=0,1,...,2^{|V|}-1$ label the elements of the powerset of $V$.\\
 The joint probability of choosing the set $X_i$ at the `$i$'th step where `$i$' ranges from $i=1,2,...,k$ is given by
\begin{align}
&\mathcal{P}^{(k)}(X_k,X_{k-1},....,X_{2},X_1)\nonumber\\
&=M^{(k)}(X_k|X_{k-1})\mathcal{P}^{(k-1)}(X_{k-1},X_{k-2},....,X_2,X_1)
\end{align}
 where $\mathcal{P}^{(k-1)}(X_{k-1},..,X_1)=\sum_{X_k}\mathcal{P}^{(k)}(X_k,X_{k-1},..,X_1)$ is the marginal distribution of the preceding $k-1$ set-valued random variables. Iterating this equation one can write the joint probability at the $k$'th step as a product of the transition matrices given an initial probability vector $\mathcal{P}^{(1)}(X_1)$ as follows:
\begin{align}
&\mathcal{P}^{(k)}(X_k,X_{k-1},...,X_1)\nonumber\\
&=M^{(k)}(X_k|X_{k-1})M^{(k-1)}(X_{k-1}|X_{k-2})\times.....\nonumber\\
&~~~~~~~~~~~~~~~~......\times M^{(2)}(X_{2}|X_{1})\mathcal{P}^{(1)}(X_1)
\end{align}
where $\mathcal{P}^{(1)}(X_1)$ is a column vector representing the probability of choosing $X_1\subset V$ in the first draw by the RQC.
In particular, for independent choices of $X_i,~i=1,..,k$ at each level of the circuit, we have that the elements of the Markov transition matrix satisfy, $M^{(i)}(X_{k}|X_{k-1})=\mathcal{P}^{(i)}(X_k)$, which implies $\mathcal{P}^{(k)}(X_k,X_{k-1},....,X_1)=\prod_{i=1}^k\mathcal{P}^{(i)}(X_i)$. The depth of this circuit is  $\prod_{i=1}^k |X_i|$. If, as we assume, the subsets $X_i$ are finite, the depth of this circuit is $O(k)$.

The associated $k-$iterated ensemble is then given by 
\begin{align}
&\mathcal{E}^{k}(\mathcal{P}^{(k)}(X_k,X_{k-1},...X_1), d\mu_i)\nonumber\\
&~~~~~~~~~~~~~~~~= \{ U_{X_k}....U_{X_2}U_{X_1}\ket{\Phi} \}_{X,U_X,\Phi}
\end{align}
where $d\mu_i, i=1,...,k$ are the measures with which the RQC chooses the unitaries to act on the corresponding chosen subsets $X_i\subset V$.

\section{Ensemble statistical moments and superoperator formulation}
Once the ensemble $\mathcal E^k$ of physical states has been constructed, we need to derive the associated ensemble for the reduced system $A$. This is 
naturally  obtained by tracing out the environment $B$, that is:
\be
\mathcal E_A^k = \{\tr _B \rho | \rho\in \mathcal E^k \}.
\ee
In order to compute the Renyi entropies for this ensemble, 
given a  density matrix $\Omega\in \mathcal E_A^k$, we will compute the trace of its  $\alpha-$power 
\be 
P^{\alpha}(\Omega)=\mathrm{Tr}_A(\Omega^{\alpha})
\ee
with   $\alpha\geq 1$. The central objects of our analysis are the statistical moments of the $P^\alpha (\Omega)$ within $\mathcal E_A^k$, for instance the average
$\overline{P^\alpha (\Omega)}^\Omega$ and higher moments. 
We want to stress that the knowledge of all the statistical moments in $\mathcal E_A^k$, is equivalent to being able to compute also all the statistical moments of observables with support on the reduced system $A$ or polynomial functions $f(\Omega)$ of arbitrary degree. One pertinent example is that knowledge of the $\alpha-$powers allow us to compute the Renyi entropies $H_{\alpha}:=\frac{1}{1-\alpha}\mathrm{log}[\mathrm{Tr}(\rho^{\alpha}_A)]$, which  are continuous w.r.t the parameter $\alpha$  \cite{ref38}. Also,  $\frac{\partial}{\partial \alpha}H_{\alpha}\leq 0$ and therefore  $H_{\alpha=1}\geq H_{\alpha=2}$. Since the VonNeumann entropy $E(\rho_A)$ can be obtained as the limit, $E(\rho_A):=\mathrm{lim}_{\alpha\to 1^+}H_{\alpha}$ we see that the 2-Renyi entropy lower bounds the VonNeumann entropy. Moreover,  for close to minimal purity the bound gets very tight \cite{ref39}.
In order to compute the statistical moments in $\mathcal E_A^k$, we will use quantum information theoretic tools, then we will introduce a superoperator formulation which will allow for a compact description of the statistical properties of the reduced system, and will show how locality of the interactions influences its entanglement properties. 
We start by recalling that  for every density matrix $\Omega$ of the reduced system $A$,
\begin{align}
P^{\alpha}&=\mathrm{Tr}_A(\Omega^{\alpha})=\mathrm{Tr}_A(\Omega^{\otimes\alpha}\tilde{T}^{(\alpha)}_A)\nonumber\\
&=\mathrm{Tr}[\omega^{\otimes\alpha}~{T}^{(\alpha)}_A]
\end{align}
where $\tilde{T}^{(\alpha)}_A:(\mathcal{H}_A)^{\bigotimes\alpha}\mapsto(\mathcal{H}_A)^{\bigotimes\alpha},\ket{i_1,i_2,....,i_{\alpha}}\mapsto\ket{i_{\alpha},i_1,i_2,....,i_{\alpha -1}}$ is the order $\alpha$ shift operator acting on the $\mathcal H_A$ subspace alone. It is thus a restriction of the operator $T^{(\alpha)}_A$ to just the $A$ subspace where $T^{(\alpha)}_A|_{(\mathcal{H}_A)^{\otimes\alpha}}=\tilde{T}^{(\alpha)}_A$ and $T^{(\alpha)}_A|_{(\mathcal{H}_B)^{\otimes\alpha}}=\mathbf{1}_{(\mathcal{H}_B)^{\otimes\alpha}}$. Here $\omega$ is the state of the total space $\Omega=\mathrm{Tr}_B[\omega]$. Note that up to a rearrangement of spaces $\tilde{T}^{(\alpha)}_A=\otimes_{i\in A}T^{(\alpha)}_i \otimes \mathbf{1}_B$. To avoid later confusion we remark at this point that we use tensored copies $\mathcal{H}^{\otimes \alpha}$ of the total Hilbert space $\mathcal{H}=\otimes_{v\in V}\mathcal{H}_v$ that is itself a tensor product of local spaces. We denote the former as rank-$\alpha$ tensor space while the latter just as a tensor product space. Note that a state in $\mathcal{H}^{\otimes \alpha}$ of the form $\psi^{\otimes\alpha}$ is symmetric under permutation operators acting on the $\alpha$ , $\mathcal{H}$ spaces. In particular the $\alpha$-tensored copy of the totally factorized state $\omega^{\otimes\alpha}$ is symmetric under $T^{(\alpha)}_A,~A\subset V$.
Starting with a fiducial (completely factorized) state $\omega = \ket{\phi}\bra{\phi} = \otimes_{i\in V}\ket{\phi_i}\bra{\phi_i}$,  the average of $P^{\alpha}$ over unitaries $U_X$ that act on the set of vertices $X$ drawn by the RQC is given by
\begin{align}
\overline{P^{\alpha}}^{U_X}&=\int d\mu(U|X)\mathrm{Tr}[(U_X\omega U_X^{\dagger})^{\otimes\alpha}~T^{(\alpha)}_A]\nonumber\\
&=\int d\mu(U|X)\mathrm{Tr}[\omega^{\otimes\alpha} (U_X^{\dagger})^{\otimes\alpha}~T^{(\alpha)}_AU_X^{\otimes\alpha}]
\label{Iop}
\end{align}
where in going to the second line above we used  cyclicity of the trace.  By choosing  the Haar measure as $d\mu$ in the integral above, we can perform integration using standard results from Representation theory of  groups \cite{ref40}-\cite{ref42}.

\subsection{General formulation}
In this section, we present a formulation in terms of superoperators in order to treat general RQCs. The key insight is given by examining the action of the  operator valued integral, $\mathcal{R}^{(\alpha)}_{X
}:\mathcal{B}(\mathcal{H}^{\otimes\alpha}_V)\to\mathcal{B}(\mathcal{H}^{\otimes\alpha}_V)$,  defined by
\be
\mathcal R^{(\alpha)}_X (\hat{O})\equiv{\int d\mu(U_X)(U_X^{\dagger})^{\otimes \alpha}\hat{O}U_X^{\otimes\alpha}}
\label{rsupop}
\ee
where as before $U_X$ are unitaries drawn with a measure $d\mu (U|X)$ acting on the set of vertices $X$ drawn by the RQC and $\hat{O}\in\mathcal{B}(\mathcal{H}^{\otimes\alpha}_V)$ is any operator on the same space. We are interested in the specific case where $\hat{O}=T^{(\alpha)}_A$. Note that $ \mathcal{R}^{(\alpha)}_X$ is not necessarily a self-dual (hermitian) superoperator and whether it is so depends on the choice of unitaries and the measure of integration.
The stochastic procedure to pick the edge $X$ is encoded by defining the  superoperator 
\be
\mathcal{R}^{(\alpha)}(\hat{O})=\sum_{X\subset V}\mathcal{P}(X)\mathcal{R}_X^{(\alpha)}(\hat{O}).
\label{stocop}
\ee
Equation (\ref{stocop}) represents the action of a RQC of depth $O(1)$. For a circuit of arbitrary depth $O(k)$, the superoperator takes the form: 
\begin{align}
&\mathcal R^{(\alpha)}_{(k)} (\hat{O})=\sum_{X_1,.....,X_k\subset V}\mathcal{P}^{(k)}(X_k,X_{k-1},...,X_1)\times\nonumber\\
&~~~~~~~~~~~~~~~\prod_{i=1}^{i=k}d\mu(U|X_i){(U_{X_k}.....U_{X_1})^{\dagger}}^{\otimes\alpha}\hat{O}(U_{X_k}.....U_{X_1})^{\otimes\alpha}
\label{kstocop}
\end{align}
 In the case of uncorrelated choices of the subsets at each step eq.(\ref{kstocop}) takes the form:
\begin{align}
&\mathcal R^{(\alpha)}_{(k)} (\hat{O})=\sum_{X_1,.....,X_k\subset V}\prod_{i=1}^k\mathcal{P}^{(i)}(X_i)\times\nonumber\\
&~~~~~~~~~~~~~~~\prod_{i=1}^{i=k}d\mu(U|X_i){(U_{X_k}.....U_{X_1})^{\dagger}}^{\otimes\alpha}\hat{O}(U_{X_k}.....U_{X_1})^{\otimes\alpha}
\label{kstocop1}
\end{align}
From eq.(\ref{kstocop1}) one can see that for the same case (of independent choices of $X_i$), $\mathcal R^{(\alpha)}_{(k)} (\hat{O})=\prod_{i=1}^k\mathcal R^{(\alpha)}_{i} (\hat{O})$. On the other hand if the choices of $X_i$ are fully correlated i.e. $\mathcal{P}^{(i)}=\mathcal{P}~,\forall~i=1,..,k$ then $\mathcal R^{(\alpha)}_{(k)} (\hat{O})=(\mathcal R^{(\alpha)})^k(\hat{O})$.\\
The superoperator formulation of the averaging procedure over the unitaries allows one to write the average $\alpha$-moment for a $k$ level RQC very compactly as (c.f. Eq.\ref{Iop}) :
\be
\overline{P^\alpha}^U_{k} = \langle \omega^{\otimes\alpha},\mathcal R^{(\alpha)}_{(k)} (T^{(\alpha)}_A)\rangle
\ee
where $\langle O_1,O_2\rangle=\mbox{Tr}(O_1^\dagger O_2)$ is the Hilbert-Schmidt inner product. 

\vspace{2in}
\subsection{Single Edge Model}
As an illustrative example, we show how the averages Eq.(\ref{Iop}) can be obtained for the simplest model consisting of two subsystems  $\mathcal H_A=\otimes_{r}\mathcal H_r$ and $\mathcal H_B=\otimes_{s}\mathcal H_s$ connected by a single edge. The graph consists of partitions A and B connected by an edge $(i,j)$ with $V=A\cup B$ and $E=\{\{i,j\}\}$, and $i\in A$, and $j\in B$. The 2-body unitary $U_X$ has support on the edge $X=\{i,j\}$, that is, the Hilbert space $\mathcal H_X = \mathcal H_i \otimes \mathcal H_j$,  where $\dim \mathcal H_i=\dim \mathcal H_j=d$. We call this single edge $X$. For sake of simplicity, we show first the calculation for $\alpha =2$ and generalize to any $\alpha$ in the following. We have 
\begin{align}
\overline{P^{\alpha=2}}^U&=\mathrm{Tr}[(\omega)^{\otimes 2}~\int dU_X(U_X^{\dagger})^{\otimes   2}T^{(2)}_AU_X^{\otimes 2}]\nonumber\\
&=\mathrm{Tr}[(\omega)^{\otimes 2}~\int dU_X(U_X^{\dagger})^{\otimes 2}T^{(2)}_iU_X^{\otimes 2}T^{(2)}_{A\backslash i}]\nonumber\\
&=\mathrm{Tr}[(\omega)^{\otimes 2}~(\frac{\mathrm{Tr}(T^{(2)}_i\Pi_+)}{d_+}\Pi_++\frac{\overbrace{\mathrm{Tr}(T^{(2)}_i\Pi_-)}^0}{d_-}\Pi_-)T^{(2)}_{A\backslash i}]\nonumber\\
&=\mathrm{Tr}[(\omega)^{\otimes 2}N_d(\openone_{i,j}+T^{(2)}_iT^{(2)}_j)T^{(2)}_{A\backslash i}]\nonumber\\
&=\mathrm{Tr}[(\omega)^{\otimes 2}N_d(T^{(2)}_{A\backslash i}+T^{(2)}_{A\cup j})]\nonumber\\
&=N_d\overbrace{\mathrm{Tr}[(\omega)^{\otimes 2}~T^{(2)}_{A\backslash i}]}^1+N_d\overbrace{\mathrm{Tr}[(\omega)^{\otimes 2}~T^{(2)}_{A\cup j}]}^1=2N_d
\label{singleedge}
\end{align}
where $N_d=\frac{d}{(d^2+1)}$ and $d_{\lambda}$ is the dimension of the Irreducible subspace labelled by $\lambda=\pm$, and  $\oplus_{\lambda}\mathcal{H}_{\lambda}=(\mathcal{H}_V)^{\otimes\alpha}$. The $\Pi_+,\Pi_-$ are the projectors onto the totally symmetric and totally anti-symmetric subspaces of $(\mathcal{H}_i\otimes\mathcal{H}_j)^{\otimes 2}$. In the third line of the above derivation the trace of the product of the  operator $T^{(2)}_i$ with the projector $\Pi_-$, is zero since $\Pi_-=\frac{\openone_{i,j}-T^{(2)}_iT^{(2)}_j}{2}\implies \mathrm{Tr}(T^{(2)}_i\Pi_-)=(1/2)\mathrm{Tr}_{i,j}[T^{(2)}_i\openone_j-\openone_iT^{(2)}_j]=0$ using the fact that the swap operators on the same subspace square to one i.e. $(T^{(2)}_{i,j})^2=\openone_{i,j}$.  In order to get to the last line in eq.(\ref{singleedge}) above we use the fact that both the traces yield $1$ since the $T^{(2)}_X,X\subset V$ operators acting on the totally symmetric state $(\omega)^{\otimes 2}$ (under permutations of the rank-2 tensor space $\mathcal{H}^{\otimes 2}$)  leave it invariant.

Note that the fact that $U_X$ has support on both $\mathcal H_A$ {\em and} $\mathcal H_B$ is crucial. Indeed, if we consider a unitary $U_A$ with support on just $A$ (and similarly on $B$), we would obtain  $\int dU \mathrm{Tr}[\omega^{\otimes 2}(U_A^{\dagger})^{\otimes 2}T^{(2)}_AU_A^{\otimes 2}]=1$ since $T^{(2)}_A\in S_2~\textrm{(symmetric~group~of~order~2)}$ is also an element of the commutant of $U^{\otimes 2}$ i.e. for $U=U_{A,B},~[U^{\otimes 2}_{A,B},T^{(2)}_A]=0$. This property will constitute the basis for the general formulation to describe general RQC. 
 
This calculation can be generalized to all $\alpha\geq3$. The average of  $P^{\alpha}$ over $U_X\in\mathcal{U}(\mathcal{H}_X=\mathcal{H}_i\otimes\mathcal{H}_j)$ is:
\begin{align}
\overline{P^{\alpha}}&=\mathrm{Tr}[\omega^{\otimes\alpha}\int dU (U^{\dagger})^{\otimes\alpha}T^{(\alpha)}_AU^{\otimes\alpha}]\nonumber\\
&=\mathrm{Tr}[\frac{\omega^{\otimes\alpha}\Pi^{\alpha}_+}{d_+}\mathrm{Tr}[\Pi^{\alpha}_+T^{(\alpha)}_A]]=\mathrm{Tr}[\frac{\omega^{\otimes\alpha}\Pi^{\alpha}_+}{d_+}]\mathrm{Tr}[\Pi^{\alpha}_+T^{(\alpha)}_A]\nonumber\\
&=\mathrm{Tr}[\frac{1}{\alpha !\binom{\alpha+d^2-1}{d^2-1}}\sum_{\sigma\in S_{\alpha}}(\sigma_i\otimes\sigma_j)(\bar{\sigma_i}\otimes\openone_j)]\nonumber\\
&=\frac{1}{\binom{\alpha+d^2-1}{d^2-1}}\frac{1}{\alpha!}(\sum_{\sigma\in S_{\alpha}}\mathrm{Tr}(\sigma_i\bar{\sigma}_i)\mathrm{Tr}(\sigma_j))\nonumber\\
&=\frac{1}{\binom{\alpha+d^2-1}{d^2-1}}\sum_{\lambda:\mathrm{IRR}~\text{of}~S_{\alpha}}\frac{m^2_{\lambda}}{d_{\lambda}}\chi_{\lambda}(\bar{\sigma}_i)\nonumber\\
\end{align}
where $\Pi^{\alpha}_+$ and $(d_+)$ are the projector and the dimension of the totally symmetric part of $\mathcal{H}_e^{\otimes\alpha}$, respectively. $m_{\lambda},d_{\lambda}$ are the multiplicities of the different Irreducible representations (IRRs) of $S_{\alpha}$ in the `$\sigma$' representation. $\chi_{\lambda}(\bar{\sigma}_i)$ is the character in those IRRs of $\bar{\sigma}_i$ which is the fixed operator that performs  the action $\mathcal{H}_i^{\otimes\alpha}\to\mathcal{H}_i^{\otimes\alpha}$ by taking a basis state $\ket{i_1,i_2,...,i_{\alpha}}\to\ket{i_{\alpha},i_{1},i_2},...,i_{\alpha-1}$ in other words the $T^{(\alpha)}_A$ operator restricted to just the $i$'th subspace. In going to the second line above we use the fact that the trace of $\omega^{\otimes\alpha}$ w.r.t. all other projectors except onto the totally symmetric one is zero. To prove  the last line we use Schur's Orthogonality theorem \cite{ref40} (or see appendix \ref{A}).

Being able to compute averages of arbitrary powers of the reduced system, we can also compute higher statistical moments. We are especially interested in the variance of the purity:
\be
Var(P^{\alpha=2})=\overline{(P^{\alpha=2})^2} - (\overline{P^{\alpha=2}})^2
\ee
For the Single Edge Model the first term is computed as follows
\begin{align}
\overline{(P^{\alpha=2})^2}&=\int [\mathrm{d}U_e]\mathrm{Tr}[\omega^{\otimes 4}~\bar{U}_e^{\otimes4}~T^{(12)}_A\otimes T^{(34)}_A~U_e^{\otimes4}]\nonumber\\
&=\mathrm{Tr}[\frac{\omega^{\otimes 4}\Pi_+}{d_+}]\mathrm{Tr}[\Pi_+(T^{(12)}_i\otimes T^{(34)}_i)]
\label{ranvariance}
\end{align}
where $d_+ =\binom{d^2+4-1}{4}=\frac{(d^2+3)(d^2+2)(d^2+1)(d^2)}{4!}$ is the dimension of the totally symmetric subspace of  $(\mathcal{H}_{i}\otimes\mathcal{H}_{j})^{\otimes 4}$ and  $\Pi_+=\frac{1}{4!}(\sum_{\sigma\in S_4}\sigma_i\otimes \sigma_j)$ is the projector onto it. In the above, only the symmetric projector is relevant because $\ket{\phi_n}\bra{\phi_n}^{\otimes 4}$ has support only on it. The operators $T^{(12)}_i,~T^{(34)}_i$ represent swaps on the first two and the last two spaces again of $(\mathcal{H}_{i}\otimes\mathcal{H}_{j})^{\otimes 4}$ i.e. in cycle notation the tensor product $T^{(12)}_i\otimes T^{(34)}_i$ represents the $(12)(34)$ element of $S_4$.\\ 
Direct calculations (see appendix \ref{B}) leads to $\overline{(P^{\alpha=2})^2} =\frac{2(2d^4+9d^2+1)}{(d^2+3)(d^2+2)(d^2+1)}$. Finally we obtain 
\be\label{var1}
Var[P^{\alpha=2}]=\frac{2(d^2-1)^2}{(d^2+3)(d^2+2)(d^2+1)^2}
\ee
If the subsystems $i,j$ are qubits, $d=2$ and we obtain $Var[P^{\alpha=2}]=.017$.

We have seen from eq.(\ref{singleedge}) that for $\alpha=2$ the superoperator on a two-vertex edge $X:=\{a,b\}$ and any subset $A\subset V$ of the graph acts like
\ba\nonumber
\mathcal R^{(2)}_X(T_A)& = &  N_d (T_{A\backslash X}+T_{A\cup X})\; X\cap A \ne\emptyset \wedge X\cap B\ne\emptyset \\
\mathcal R^{(2)}_X(T_A)& = &  T_A \qquad \text{otherwise}
\label{algebra}
\ea
From this key relation we see that the $X-$supported superoperator takes a permutation operator $T_A$ with support in $A$ and yields two permutation operators with support in $A\backslash X$ and $A\cup X$, respectively, {\em when $X$ is across the boundary between $A$ and $B$}, otherwise is trivial. This action has two implications: (i) a reduction of purity (because the number $N_d<1$ appears). Therefore only when $\mathcal R_X$ is supported across the boundary it can have a non trivial effect, and (ii) a propagation in the bulks of both $A$ and $B$ at a distance given by the diameter of $X$. When iterating with $k$, this implies a propagation of entanglement in the bulk that scales with $k$, as we will see in the concrete models described in the next sections. In the case of general hyper graphs, the edges may contain any number of vertices.\\
In the following, we will exploit the power of this formulation in order to study models defined on non trivial graphs. In this way, we will be able to show typicality of area law and volume law for the entanglement of a subsystem. 

\section{Random Edge Model}
\subsection{Random Edge Model for a general graph}
In this model, we consider a graph $(V,E)$ where the set of edges $E\subset V^2$, that is, a usual graph.  The RQC draws an edge $X\in E$ according to the flat unit normalized measure $\mathcal{P}:X\subset E\to \mathcal{P}(X)\in[0,1]$, with $\mathcal{P}(X)=\frac{1}{|E|}$. Conditioned to the extraction of the edge $X$, a unitary with support on $X$ is drawn with the Haar measure 
 $d\mu_{Haar}(U|X)$.
In this way, we obtain the ensemble  
\be
\mathcal E (\mathcal{P}=\frac{1}{|E|}, d\mu_{Haar}) = \{ U_X\ket{\phi} \}_{X,U_X,\phi}
\ee
where as usual $\phi$ is the completely factorized fiducial state. This ensemble is obtained by extracting RQC of depth $\le \max_{X\in E}|X|=O(1)$.
If we denote by $\partial A \subset E$ those edges that go across the boundary, that, is those that have non-null intersection with both $A$ and $B$, the average purity is given by 
\begin{align}
\overline{P^{\alpha=2}}&=\sum_{X\in E\backslash \partial A}\mathcal{P}(X)\mathrm{Tr}[\omega^{\otimes 2}\int d[U_X](U^{\dagger}_X)^{\otimes 2}T_A (U_X)^{\otimes 2}]\nonumber\\&~~~~~+\sum_{X\in \partial A}\mathcal{P}(X)\mathrm{Tr}[\omega^{\otimes 2}\int dU_X(U^{\dagger}_X)^{\otimes 2}T_A(U_X)^{\otimes 2}]\nonumber\\
&=\sum_{X\in E\backslash \partial A}\mathcal{P}(X) \times  1 \nonumber\\&~~~~~+ \sum_{X\in \partial A}\mathcal{P}(X)[N_d\mathrm{Tr}[\omega^{\otimes 2}T_{A\cup X}] + N_d\mathrm{Tr}[\omega^{\otimes 2}T_{A\backslash X}]]\nonumber\\
\label{REM1}
&=(1-q)+2qN_d
\end{align}
where $q=\sum_{X\in \partial A}\mathcal{P}(X)=\frac{\partial A}{|E|}$ is the net probability of a boundary node of $A$ interacting with one in $B$. Since local unitaries completely \emph{internal} to $A$ or $B$  i.e. with no support across the bipartition cannot affect the purity we get a 1 for the integral in the first term on the R.H.S..  Note that this averaging procedure over the edge unitary executed once leads to an algebra of the $T_X$ operators where with probability $q=\frac{\partial A}{|E|}$ the RQC generates an equal superposition of $T_{A\cup e}$ and $T_{A\backslash e}$ ($e$ being the edge on which the unitary acts) and with the complement of the probability $(1-q)$ it leaves the operator $T_A$ invariant. 
At this point, we are ready to show how to compute the variance of the purity $P^{\alpha=2}$ in the above ensemble.\\
 \\
\indent By using the technique leading to Eq.(\ref{var1}) and the probability distribution $\mathcal{P}(X)$ of the REM, We obtain (with $I=\overline{(P^{\alpha=2})^2}$ defined in eq.(\ref{ranvariance})):
\begin{align}
Var[P^{\alpha=2}_{REM}]&=(1-q+q\overline{(P^{\alpha=2})^2})-(1-q+q2N_d)^2\nonumber\\
&\approx q(1+I-4N_d),~~~~(\text{for}~q<<1)\nonumber\\
&=\frac{|\partial A|}{|E|}(1\!-\!\frac{4d^5\!-4d^4\!+20d^3\!-18d^2\!+24d\!-\!2}{(d^2\!+\!3)(d^2\!+\!2)(d^2\!+\!1)})
\end{align}
This is the variance obtained by the distribution obtained for a single edge extraction. We want now to compute the statistical momenta for the ensemble in which we extract the edges a number $k$ of times. We iterate the procedure by using sequential independent identical RQCs and build the $k-$iterated ensemble $\mathcal{E}^{(k)}$. Let us consider unitaries of the form $U=\prod_{i=1}^k U_{X_{i}}$ where the $X_i$'s and $U_{X_i}$'s are drawn according to the probability distributions $\prod_{i=1}^k \mathcal{P}(X_i)$ and $\prod_{i=1}^k d\mu(U_{X_i}$ respectively. If we now assume that the degree of each vertex is $o(|\partial A|)$ which implies that the boundary length changes negligibly due to the algebra of eq.(\ref{REM1}) i.e we may take $|\partial A\cup e|\approx|\partial A\backslash e|\approx|\partial A|$ then for the second iteration of the procedure  we find that the average purity is given by :
\begin{align}
\overline{P^{\alpha=2}_{k=2}}&=\sum_{\substack{X\in E \backslash \partial A \\ X'\in E\backslash \partial A}}\mathcal{P}(X)[\mathcal{P}(X') T_A + \mathcal{P}(X')N_d(T_{A\cup e'} + T_{A\backslash e'})]\nonumber\\
&+\sum_{\substack{X\in \partial A\\ X'\in E\backslash \partial(A\cup e)}}\mathcal{P}(X)N_d[\mathcal{P}(X')T_{A\cup e}+\mathcal{P}(X')T_{A\backslash e}]\nonumber\\
&+\sum_{\substack{X\in \partial(A\cup e)\\ X'\in \partial(A\backslash e) }}\mathcal{P}(X')N_d[N_d(T_{A\cup e \cup e'}+T_{A\cup e \backslash e'})\nonumber\\
&~~~~~~~~~~~~~~~~+\mathcal{P}(X')N_d(T_{A\backslash e \cup e'}+T_{A\backslash e\backslash e'})]\nonumber\\
&~~~~~\sim ((1-q)+2qN_d)^2
\end{align}
It can be shown that under the same assumption (of negligible change of boundary length) the purity for $k$ iterations goes as $\overline{P^{\alpha=2}_k}=(1-q(1-2N_d))^k$. One can understand the physical content of these calculations by considering the thermodynamic limit of large $|E|>>|\partial A|\implies q$ small.In this limit the average 2-Renyi Entropy $\overline{H}_2$ can be lower bounded (using concavity of the log function) by the logarithm (base 2) of the average purity i.e.:
\begin{align}
\overline{H}_2&:=-\overline{\mathrm{log}P^{\alpha=2}_k}\geq-\mathrm{log}\overline{P^{\alpha=2}_k}\nonumber\\
\implies &\overline{H}_2\geq -k\mathrm{log}(1-q(1-2N_d))\sim qk(1-2N_d)
\label{renyi}
\end{align}
The number $k$ of iterations corresponds to  the time in our scheme (and in the circuit model), so Eq.(\ref{renyi}) implies linear increase of the entropy with time.It also implies that the entropy is proportional to the boundary of the region $|\partial A|$. For this model,  the $\alpha-$Renyi entropy turns out to have the form $H_{\alpha}\geq \frac{1}{1-\alpha}\mathrm{log}[\mathrm{Tr}[\overline{P^{\alpha}}]]\propto |\partial A|$ which can be seen by considering $\overline{P^{(\alpha)}}=(1-q)+q\mathrm{Tr}[\omega^{\otimes\alpha}\int d[U_X](U^{\dagger}_X)^{\otimes 2}T^{(\alpha)}_A (U_X)^{\otimes 2}]
=1+q(C(\alpha,d)-1)$ where $C(\alpha,d)=\frac{\mathrm{Tr}[T^{(\alpha)}\Pi^+]}{\binom{\alpha+d^2-1}{d^2-1}}$ is a group theoretic factor less than 1 with $\Pi^+$ the projector onto the totally symmetric subspace of $\mathcal{H}^{\otimes\alpha}$.

All of the above calculations can be reformulated in terms of the expectation value of the stochastic Hermitian superoperator $\mathcal{R}=\sum_{X\in V}\mathcal{P}(X)\mathcal{R}_X=\sum_{X\in V}\mathcal{P}(X)\int dU_X(U^{\dagger})^{\otimes 2}T_XU^{\otimes 2}$. 
In the next sections, we show that the asymptotic analysis of $\mathcal{R}$ for this case leads to a asymptotic purity of the form as in Eq.(\ref{renyi}).

\subsection{Random Edge Model for the complete graph $K_N$}

Let us now consider the case in which the graph of the REM is the complete graph $K_N$, that is, a graph in which every two vertices are connected by an edge: $E=V\times V$. The number of edges is of course $|E| = |V^2|= \binom{N}{2}$. As before,  the RQC draws an edge $X$ according to flat unit normalized measure $\mathcal{P}:X\in E\to \mathcal{P}(X)\in[0,1]$, with $\mathcal{P}(X)=\frac{1}{|E|}$. Conditioned to the extraction of the edge $X$, a unitary with support on $X$ is drawn with the Haar measure  $d\mu_{Haar}(U|X)$. The analysis of this case highlights the utility of the superoperator formalism. We introduce a bipartition in the system by $V=A\cup B$ with $A\cap B=\emptyset$, with cardinalities $|A|=N_A$ and $|B|=N_B$, and of course $N=N_A+N_B$.
The average purity after the $k$-th iteration is given by: 
\be
\overline{P^{\alpha=2}_k}(T_A)=\mathrm{Tr}[\omega^{\otimes 2}\mathcal{R}_k^{(2)}(T^{\otimes 2}_A)]
\ee
 To see what form $\mathcal R$ takes we have to see that after each extraction, if the extracted edge is not straddling the bipartition then the superoperator has trivial action. Therefore, the probability $q(A)$ that $\mathcal R(T_A)$ has a trivial action is given by the probability of drawing an edge completely inside $A$ or completely inside $B= V\backslash A$, that is, $q(A)=\frac{\binom{N_A}{2}+\binom{N-N_A}{2}}{\binom{N}{2}}$. Otherwise, $\mathcal R(T_A)$ has a non trivial action:
\begin{align}
\mathcal{R}^{(2)}(T_A)=q(A)T_A+\frac{N_d}{|E|}\{N_B\sum_{i\in A}T_{A\backslash i}+N_A\sum_{j\in B}T_{A\cup j}\}
\label{supop}
\end{align}
Notice that the superoperator $\mathcal{R}^{(2)}$ acts on the space of shift operators $T_A$. That is, for any subset $A\subset V$, $\mathcal{R}^{(2)}(T_A)$ is an operator in Hilbert space $\mathcal H^{\otimes 2}$. Therefore, the matrix elements of $\mathcal{R}^{(2)}$ in the space of the shift operators are given by 
\be
\mathcal{R}^{(2)}_{B,A}=\bra{T_B}\mathcal{R}^{(2)}\ket{T_A}
\ee
Notice now that the $k-$th iteration of the superoperator corresponds to matrix multiplication, that is, 
\be
(\mathcal{R}^{(2)}_k)_{B,A}=\bra{T_B}(\mathcal{R}^{(2)})^k\ket{T_A}
\ee
In order to find a convenient expression for $\mathcal R$ for this model, we show how to map the superoperator $\mathcal{R}(T_A)$ in Eq.(\ref{supop}) acting on the swap operators to a spin operator of the $2^N$-dim Hilbert space (we are assuming now we are dealing with qubits). First of all, consider the mapping between subsets of $V$ to a pure vector which is a member of the computational basis in the $2^N$-dim Hilbert space of the qubits:
\ba\nonumber\label{compbasis}
K :A\subset V\to\otimes_{i\in V}\ket{\chi_A(i)}&=&\ket{\psi_A} \\
 \mbox{with} ~\chi_A(i)&=&\begin{cases}0~~i\in A\\1~~i\notin A\end{cases}
\ea
Here, $\chi_A(i)$ is the indicator function for the node $i\in V$. By considering all the subsets $X$ in the power set $\mathcal P(V)$ it is obvious that the states $\ket{\psi_X},~X\subset V$ form a complete orthonormal basis for the $2^{N}$ dimensional Hilbert space of $N$ spin 1/2 particles. We will call this space the abstract qubit space and operators in this space are denoted with a hat on top. On this space let us introduce the total spin operators $\hat{S}^{\alpha}=\frac{1}{2}\sum_{i\in V}\sigma^{\alpha}_i$ where $\sigma^{\alpha}_i$ are the Pauli matrices acting at the i'th spin. Note that then:
\begin{align}
\hat{S}^z\ket{\psi_A}&=\frac{1}{2}(N_A-N_B)\ket{\psi_A}
\end{align}
which allows us to promote the numbers $N_A,N_B$ in Eq.(\ref{supop}) to operators $N_A\to\hat{N}_A = (\hat{S}^z+N/2), N_B\to \hat{N}_B= (N/2-\hat{S}^z)$, in the sense that 
$N_A = \langle\psi_A| \hat{N}_A | \psi_A\rangle$ and similarly for $\hat{N}_B$. With these definitions, we have $q_A=1-\frac{N_AN_B}{N(N-1)}=1-\frac{1}{|E|}[(N/2)^2-(\hat{S}^z)^2]$. Similarly we now define the total raising $\hat{S}^+=\hat{S}^x+i\hat{S}^y$ and lowering operators $\hat{S}^-=\hat{S}^x-i\hat{S}^y$. Then we obtain
\begin{align}
\hat{S}^-\ket{\psi_A}&=\sum_{j\in B}\ket{\psi_{A\cup j}}\\
\hat{S}^+\ket{\psi_A}&=\sum_{i\in A}\ket{\psi_{A\backslash i}}
\end{align}
\begin{figure}
 \centering
 \includegraphics[]{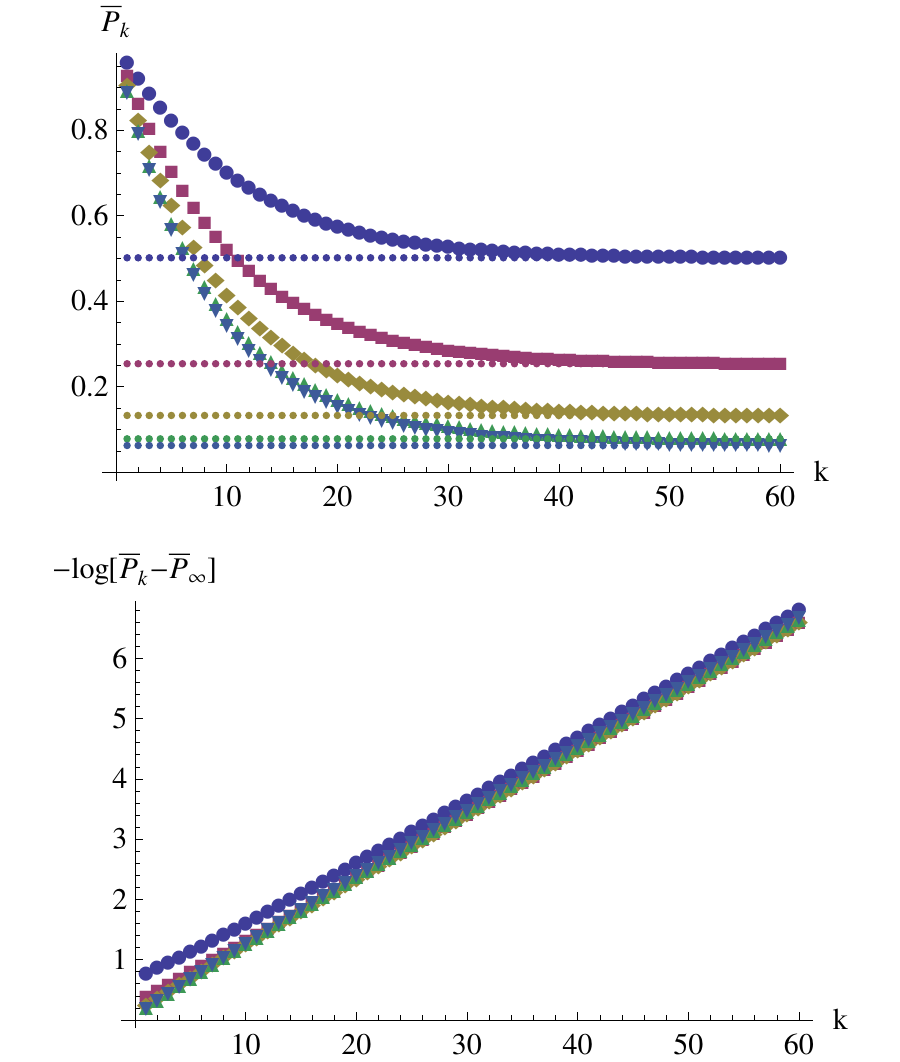}
 \caption{[\romannumeral 1] (left panel) Purity as given by Eq.(\ref{purity}) with fixed total size $N=10$ and different subsystem sizes $N_A=1 (\text{blue circles}),L_N=2 (\text{purple squares}),N_A=3 (\text{yellow rhombi}),N_A=4(\text{green upright triangles}),N_A=5(\text{dark blue inverted triangles})$ [\romannumeral 2] (right panel) logarithm of the difference between purity at the $k$'th step and the asymptotic purity with $N=10$ and same color codes for $N_A$ as in [\romannumeral 1].}
  \label{pur10}
\end{figure} 
Now we can define the (hatted) operator $\mathcal{\hat{R}}^{(2)}$ acting on the abstract qubit space as:
\begin{align}
\mathcal{\hat{R}}^{(2)}&:= (1-\frac{1}{|E|}[(N/2)^2-(\hat{S}^z)^2])\nonumber\\
&+\frac{N_d}{|E|}[\hat{S}^+(N/2-\hat{S}^z)+\hat{S}^-(N/2+\hat{S}^z)]
\label{absop}
\end{align}
$\mathcal{\hat{R}}^{(2)}$ is a non-Hermitian operator in the abstract qubit space spanned by the vectors $\ket{\psi_A}$. In this space, its matrix elements read
$ (\hat{\mathcal R}^{(2)})_{B,A}=\bra{\psi_B} \hat{\mathcal R}^{(2)}\ket{\psi_A}$. 
We now show that Iterations of the RQC protocol in the original space correspond to iterations of the operator (\ref{absop}) in the abstract space. 
To this aim, an important observation is that in the original space the trace of the powers of the superoperator $\mathcal{R}$ is taken w.r.t. $\omega^{\otimes 2}$ which belongs to the totally symmetric subspace of $\mathcal{H}^{\otimes 2}$. We can finally relate the average purity of the subsystem $A$ to a sum over matrix elements of the matrix $(\hat{\mathcal R}^{(2)})^k$:
\begin{align}
\overline{P^{\alpha=2}}_k (A) &=\mathrm{Tr}[\omega^{\otimes 2} ~\mathcal{R}_k^{(2)}(T_A^{\otimes 2})]\nonumber\\
&=\langle\omega^{\otimes 2},\sum_{B\subset V}(\mathcal{R}^{(2)}_k)_{B,A}T_B\rangle\nonumber\\
&=\sum_{B\subset V}(\mathcal{R}^k)_{B,A}\overbrace{\langle\omega^{\otimes 2},T_B\rangle}^1\nonumber\\
&=\sum_{B\subset V}\bra{\psi_B}(\hat{\mathcal R}^{(2)})^k\ket{\psi_A}\nonumber\\
&=\sum_{B\subset V}((\hat{\mathcal R}^{(2)})^k)_{B,A}
\label{newop}
\end{align}
and thus from Eq.(\ref{newop}) one can see that in the abstract qubit space 
\be
\overline{P^{\alpha=2}}_k=(\bra{0}+\bra{1})^{\otimes N}(\mathcal{\hat{R}}^{(2)})^k\ket{\psi_A}
\label{smsb}
\ee
because  $\sum_{B\subset V}\ket{\psi_B}=(\ket{0}+\ket{1})^{\otimes N}\equiv \ket{\psi_{symm}}$.
Notice that the state $\ket{\psi_{symm}}$ is invariant under the action of the projector $\Pi_{S}=\frac{1}{N!}\sum_{\sigma\in S_N}\sigma$ onto the symmetric subspace of the abstract qubit space, where $\sigma\in S_N$ are the elements of the symmetric group on $N$ labels. With this in mind and the fact that Total spin operators commute with $\Pi_{S}$ we can write Eq.(\ref{smsb}) as:
\begin{align}
\overline{P^{\alpha=2}}_k&=\bra{\psi_{symm}}\Pi_{S}(\mathcal{\hat{R}}^{(2)})^k\Pi_{S}\ket{\psi_A}
\label{pur}
\end{align}
Note that $\Pi_{S}\ket{\psi_A}\in \mathcal{H}_{J=J_{max}}$ i.e. $\Pi_{S}$ is the projector onto the highest total spin subspace. In our case then it projects onto the $J=N/2$ subspace. Since $[(\mathcal{\hat{R}}^{(2)})^k,\Pi_{S}]=0$ we focus on just the $\mathcal{H}_{J=J_{max}}$ subspace. However $\Pi_S\ket{\psi_A}$ is not normalized. Normalization involves some algebra to relate states in the tensor product basis for the abstract qubit space to the total spin basis: suppose then that the z-component of spin for some state in the tensor product basis of L spins is m. Then $N_{up}+N_{down}=N,\frac{1}{2}(N_{up}-N_{down})=m\implies N_{up}=(N/2+m)$ thus the number of distinct states in the tensor product basis with m as their z-component is $\binom{N}{N_{up}}=\binom{N}{N/2+m}$. The state in the total spin basis in the $\mathcal{H}_{J=J_{max}}$ subspace that has the same z-component is a symmetric combination of these distinct states with appropriate normalization. It turns out that:
\begin{align}
\Pi_S\ket{m}=\frac{1}{\sqrt{\binom{N}{N/2+m}}}\ket{N/2,m}
\end{align}
With this in mind we see that $\Pi_S\ket{\psi_A}=\frac{1}{\sqrt{\binom{N}{N_A}}}\ket{N/2,\frac{1}{2}(2N_A-N)}$. Inserting this and the identity in the $\mathcal{H}_{J=J_{max}}$  subspace into Eq.(\ref{pur}) we obtain:
\begin{align}
\overline{P^{\alpha=2}}_k&=\frac{1}{\sqrt{\binom{N}{N_A}}}\sum_{m=-N/2}^{N/2}\bra{\psi_{symm}}\Pi_S\ket{L/2,m}\nonumber\\
&~~~~~~~~~~~~~~~\times\bra{N/2,m}(\mathcal{\hat{R}}^{(2)})^k\ket{N/2,\frac{1}{2}(2N_A-N)}\nonumber\\
&=C(N,N_A)\sum_{m=-N/2}^{N/2}\sqrt{\binom{N}{m+N/2}}(\mathcal{\hat{R}}^{(2)})^k_{m,\frac{1}{2}(2N_A-N)}\nonumber\\
\label{purity}
\end{align}
where $C(N,N_A)=\sqrt{\frac{N_A!(N-N_A)!}{N!}}$ and the matrix element of the $k$ iterated superoperator are $(\mathcal{\hat{R}}^{(2)})^k_{m,\frac{1}{2}(2N_A-N)}=\bra{\frac{N}{2},m}(\mathcal{\hat{R}}^{(2)})^k\ket{\frac{N}{2},\frac{1}{2}(2N_A-N)}$. \\
\indent Note that Eq.(\ref{purity}) expresses purity dynamics of a $d^{N}$-dimensional system in terms of dynamics in an exponentially smaller $(N+1)$-dimensional space. From eq.(\ref{absop}) one can see that for $N_A=0, \mathcal{\hat{R}}^{(2)}\ket{\psi_A}=1$ as well as for $N_A=N, \mathcal{\hat{R}}^{(2)}\ket{\psi_A}=1$, all other eigenvalues of the operator being less than 1. If one assumes that these are the only two fixed points of the operator then asymptotically the non-zero eigenspace is spanned by the symmetric and antisymmetric combinations $\frac{\ket{\psi_0}+\ket{\psi_N}}{\sqrt{2}},\frac{\ket{\psi_0}-\ket{\psi_N}}{\sqrt{2}}$ with only the symmetric combination contributing to the purity which reaches the value 
\be\label{asymptotic}
\overline{P^{\alpha=2}}_{k\to\infty}=\frac{d^{2N-N_A}+d^{N+N_A}}{d^N(d^N+1)}
\ee 
The typical behavior of purity is shown in fig.(\ref{pur10}).
\begin{figure}
 \centering
 \includegraphics[]{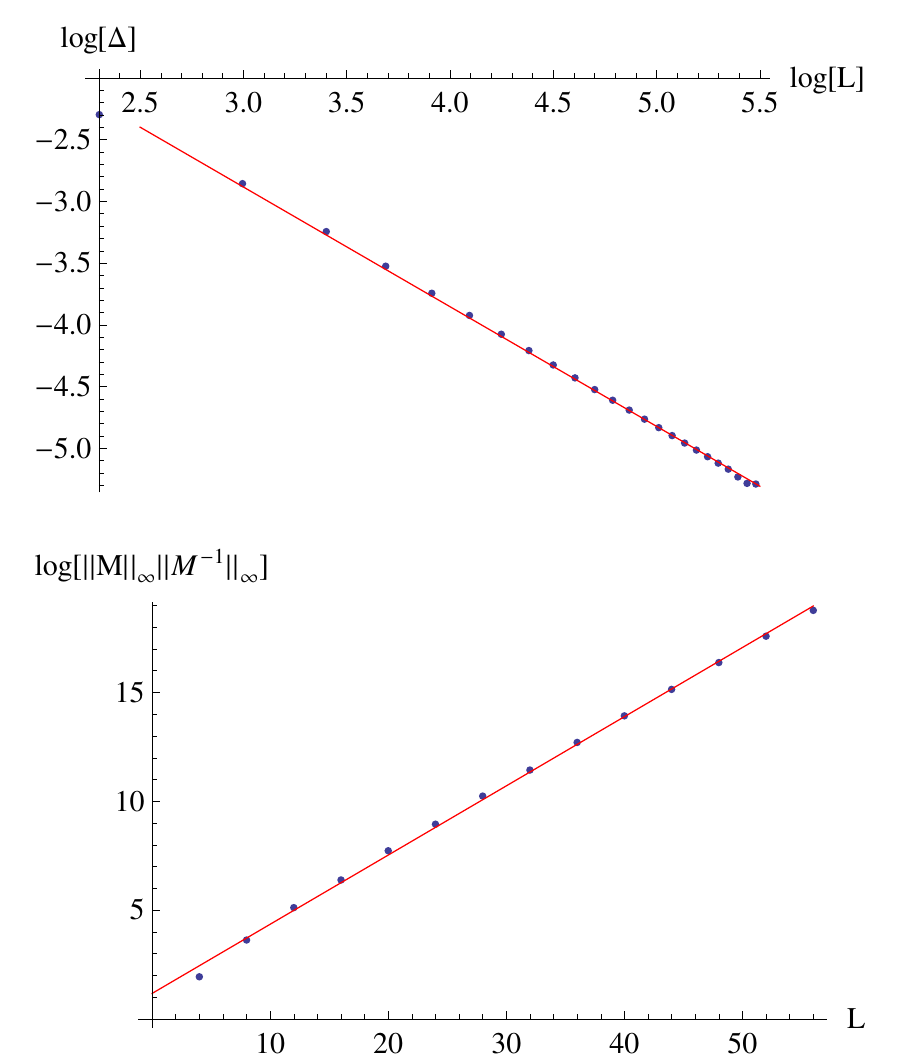}
 \caption{[\romannumeral 1] (Top panel) Scaling behavior of gap($\Delta$) for the superoperator of the Fully connected RQC. The best fit line(red) gives us the equation: $\log[\Delta]=-.969395\log[L]+.0246366$ [\romannumeral 2](Bottom panel) Scaling behavior of the logarithm of the product of operator norms for the similarity transform matrix $M$ and it's Inverse. The best fit line(red) gives $\log[||M||_{\infty}||M^{-1}||_{\infty}]=1.17508 + 0.318L$}
  \label{boundfig}
\end{figure} 
An obviously important  question is : How fast in $k$ does the protocol take the purity to within $\epsilon>0$ of the asymptotic value ? The answer to this question can be related to the gap of the matrix  $\mathcal{\hat{R}}^{(2)}$ in the $\mathcal{H}_{J=J_{max}}$ subspace. From Eq.(\ref{absop}) it is easy to see that the matrix representation $R$ has elements:
\begin{align}
&R_{p,q}\equiv\bra{\frac{N}{2},p}\mathcal{\hat{R}}^{(2)}\ket{\frac{N}{2},q}\nonumber\\
&=\delta_{p,q}f(p)+\frac{N_d}{|E|}[(\frac{N}{2}+q)\sqrt{\frac{N}{2}(\frac{N}{2}+1)-p(p-1)}\delta_{p,q+1}\nonumber\\
&~~~~~~~~~+(\frac{N}{2}-q)\sqrt{\frac{N}{2}(\frac{N}{2}+1)-p(p+1)}\delta_{p,q-1}]
\label{element}
\end{align}
which implies that the above matrix $R$ is tri-diagonal and satisfies the condition $R_{k,k+1}\cdot R_{k+1,k}\geq 0$. Such a matrix is similar to a Hermitian matrix. One can then diagonalize the Hermitian matrix and asymptotics can be calculated by the power method of eigenvalues for the derived Hermitian Matrix. We then have that $SRS^{-1}=H\implies \exists~U~s.t.~U^{\dagger}HU=D\implies U^{\dagger}SRS^{-1}U=D $ where $D,U,S$ are diagonal, Unitary and Invertible matrices respectively. Using the spectral resolution of $D$ one can write,$ USRS^{-1}U^{\dagger}=\sum_{i}\lambda_i\ket{\lambda_i}\bra{\lambda_i} \implies R^k=S^{-1}U^{\dagger}\sum_{i}\lambda_i^k\ket{\lambda_i}\bra{\lambda_i}US$
Then with $M=U^{\dagger}S$ we obtain from Eq.(\ref{pur}) for the case $|A|=N/2$:
\begin{align}
\overline{P^{\alpha=2}}_k&={C}(N,N_A)\sum_{\alpha=1}^{N+1}\lambda_{\alpha}^k\bra{\psi^+}M^{-1}\ket{\lambda_{\alpha}}\bra{\lambda_{\alpha}}M\ket{\psi^0}\nonumber\\
&=\frac{2^{N/2}}{\sqrt{\binom{N}{N/2}}}\sum_{\alpha=1}^{N+1}\lambda_{\alpha}^ka_{\alpha}
\end{align}
where $\ket{\psi^+}$ is the normalized $\ket{\psi_{symm}}$  and $\ket{\psi^0}=\ket{J=N/2,m=0}$ are unit normalized vectors and $a_{\alpha}=|\bra{\psi^+}M\ket{\lambda_{\alpha}}\bra{\lambda_{\alpha}}M^{-1}\ket{\psi^0}|$. Assuming that the eigenvalues of $R$ are so arranged that $\lambda_1=\lambda_2=1$ and the rest are arranged in a non-increasing order we find:
\begin{align}
&|\overline{P^{\alpha=2}_k}-\overline{P^{\alpha=2}_{\infty}}|={C}(N,N_A)|\sum_{\alpha=3}^{L+1}\lambda_{\alpha}^ka_{\alpha}|\nonumber\\
&~~~~~~~~\leq {C(N,N_A)}\sum_{\alpha=3}^{L+1}|\lambda_{\alpha}|^k|\bra{\psi^+}M\ket{\lambda_{\alpha}}\bra{\lambda_{\alpha}}M^{-1}\ket{\psi^0}|\nonumber\\
&~~~~~~~~\leq {C(N,N_A)}|\lambda_3|^k\sum_{\alpha=3}^{L+1}|\bra{\psi^+}M\ket{\lambda_\alpha}||\bra{\lambda_{\alpha}}M^{-1}\ket{\psi^0}|\nonumber\\
&~~~~~~~~\leq {C(N,N_A)}|\lambda_3|^k \nonumber \\
&~~~~~~~~\times\sqrt{\sum_{\alpha=3}^{L+1}|\bra{\psi^+}M\ket{\lambda_{\alpha}}|^2}\sqrt{\sum_{\alpha=3}^{L+1}|\bra{\lambda_{\alpha}}M^{-1}\ket{\psi^0}|^2}\nonumber\\
&~~~~~~~~\leq {C(N,N_A)} \lambda_3^k||M||_{\infty}||M^{-1}||_{\infty}
\label{kk}
\end{align}
We require the R.H.S of the above expression to be less than $\epsilon$. Therefore, taking the logarithm of the inequality ${C(N,N_A)} \lambda_3^k||M||_{\infty}||M^{-1}||_{\infty}\leq \epsilon$ we obtain (using $\lambda_3=1-\Delta$) :
\begin{align}
k&\geq\frac{\log{{C(N,N_A)}}+\log{||M||_{\infty}}||M^{-1}||_{\infty}+\log{1/\epsilon}}{\log{\frac{1}{1-\Delta}}}
\label{kkbound}
\end{align}
Since $\log{\frac{1}{1-\Delta}}\geq \Delta$, requiring 
\begin{align}
k\geq \frac{\log{{C(N,N_A)}}+\log{||M||_{\infty}}||M^{-1}||_{\infty}+\log{1/\epsilon}}{\Delta}
\label{kbound}
\end{align}
makes sure that inequality Eq.(\ref{kkbound}) is also fulfilled.
Indeed the bound in Eq.(\ref{kbound}) is a rather weak lower bound as for all graph sizes $N$ that we studied numerically the asymptotic values for any size of the subsystem $N_A<N$ were reached much before the above bound.

The main difficulty in estimating $k_{min}(\epsilon,N,N_A)$, i.e. the minimum number $k$ of iterations required to reach within $\epsilon$ accuracy of the asymptotic value resides in the calculation of the operator norms of the similarity transform matrices $M$, its inverse $M^{-1}$ and the gap $\Delta$. This amounts to diagonalizing the non-Hermitian matrix $R$ in the maximal total spin subspace. However this is an exponentially reduced problem of diagonalization in a $(N+1)$-Dim space compared to a $d^{N}$ dimensional one. Numerical study shows that the gap $\Delta$ of $R$ to have an algebraic dependence on $N$ as shown in Fig.(\ref{boundfig}). From the least-squares best fit line we can evaluate 
\be\label{fit}
\Delta=e^{.025}/N^{.97}\approx\frac{1.025}{N}
\ee 
Moreover, from the lower panel of the same figure we find a linear dependence of the logarithm of the product of operator norms of the matrices $M,M^{-1}$ i.e. 
\be\label{fit2}
\log[||M||_{\infty}||M^{-1}||_{\infty}]=1.17508 + 0.318N
\ee
Thus $\log{\mathcal{C(L)}}\approx_{L\to\infty}\alpha_1\log{L},~\log[||M||_{\infty}||M^{-1}||_{\infty}]\approx\alpha_2L+\beta, \frac{1}{\Delta}\approx\alpha_3L$ where $\alpha_1,\alpha_2,\alpha_3,\beta=O(1)$ we finally obtain the scaling 
\be
k\geq (\alpha_1\log{N}+\alpha_2N+\beta)\alpha_3N+\log{(1/\epsilon)}\alpha_3N=O(N^2)
\ee


\section{Contiguous Edge Model}
This model intends to  mimic evolution of a multi-partite system under a local (time-dependent) Hamiltonian.  Here, \emph{all} the edges of graph $\Gamma$ are acted on by the RQC with 2-local unitaries in some particular order `$\sigma$' - denoting an ordered sequence of edges. A RQC of depth $O(N)$ picks this ordering of nodes such that finally all nodes in the graph are acted on by nearest neighbour unitaries. One pass through such a circuit is called a cycle. This procedure is then iterated through $n_c$ cycles. The composite unitary of a cycle  is the $\sigma-$ordered product of unitaries with support on nearest neighbours according to the graph $\Gamma$. The ensemble is constructed by considering all the possible orderings $\sigma$.

While our general formulation of this model extends to graphs with any geometry in any number of dimensions we present a detailed analysis of 1-D graphs and the 2-D square lattice.

\begin{figure}
 \centering
 \includegraphics[width=\columnwidth]{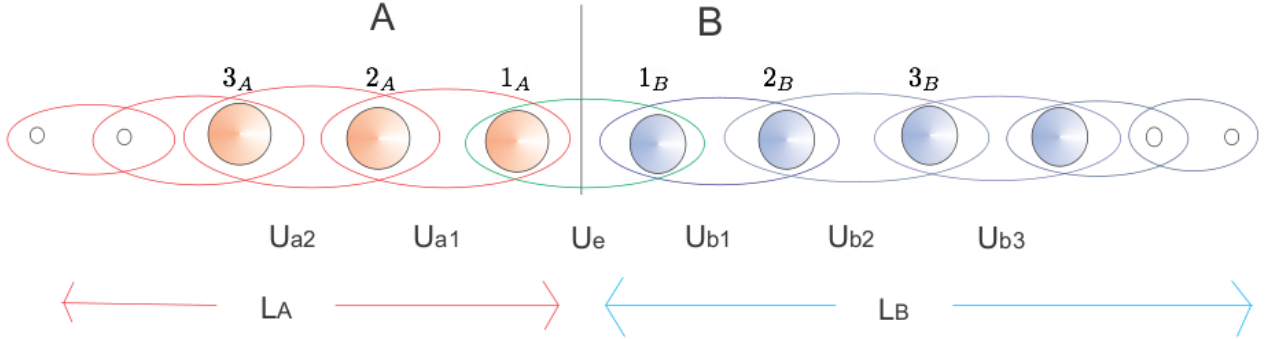}
 \caption{A bipartite $(A,B)$ spin chain of length $L=L_A+L_B$ with nearest-neighbor qubits interacting via $2-$qubit gates (ellypses). The edge $e$ is the one that straddles the two partitions. The gates are numbered by the subscript $x_i$ where $x=A,B$ denotes the two halves of the chain and $i$ the distance from the boundary of the two partitions.}
  \label{linchain}
\end{figure} 
\subsection{Contiguous Edge Model on the Linear Chain}
Let us start with the 1-D graph of Fig.(\ref{linchain}), with $N=L$, and introduce a bipartition into subsystems $A$ and $B$ of lengths $L_A,L_B$. The sites in $A$ on the left of the boundary are labeled by $i_A=1_A,...,L_A$ increasing towards the left, while the sites in $B$ are labeled by $i_B=1_B,...,L_B$ increasing towards the right. The edges on the chain are labeled by $a_i=\langle(i+1)_A,i_A\rangle$, and the edges in $B$ are labeled by $b_i=\langle i_B,(i+1)_B$. The RQC chooses an ordering among all the edges in the chain, and then on each edge acts with a  Haar distributed 2-qudit unitary operator in the given order. The ensemble therefore contains   all possible permutations of the list of edges. For $n_c=O(1)$, the average purity depends strongly on the order in which the RQC chooses the edges, while when $n_c$  exceeds the subsystem size, ordering does not really count, as we shall see in the following. In any case, we can consider the orderings that give the two extreme situations, that is, the minimum and maximum decrease of purity.  i.e. maximum entangling power which we call the best sequence and the one that corresponds to the minimal decrease of purity or minimum entangling power is termed the worst sequence. Let $U_{\sigma}$ denote the ordered product of 2-qudit unitaries over all the edges in $E$ with the order given by the permutation $\sigma$ i.e. $U_\sigma = U_{\sigma(e_1)}  \ldots U_{\sigma(e_{|E|})}$. The ensemble and its measure are then given by 
\begin{align}
\mathcal{E}_{(\sigma)} (\Gamma)&=\{U_\sigma\ket{\Phi}\}_{U}\nonumber\\
d\mu(U)&= \delta (U-U_\sigma)\prod_{e\in E} d\mu_{Haar}(U_e)
\label{CEM}
\end{align}
We can now see why sequences of unitaries corresponding to different permutations yield different purities. We start showing which sequence produces the greatest decrease of purity, that we dub {\em best case}. Consider the sequence $U_{best}=U_eU_AU_B$ where $U_e$ is the unitary straddling the edge i.e. acting on $1_A$ and $1_B$ and $U_A=U_{a_1}U_{a_2}...U_{a_{L_A-1}}$ is the \emph{internal} structure of $U_A$ where $U_{a_1}$ means a 2-qudit unitary with support on the qudits the nearest of which is at a distance of 1 lattice spacing from the boundary. Note that the internal structure of the unitary is non-deformable since $[U_{a_i},U_{a_{i+1}}]\neq 0$ as they share a node. With the same convention the internal structure of $U_B=U_{b_1}U_{b_2}...U_{b_{L_B-1}}$. Physically this corresponds to the RQC choosing and applying all possible unitaries on the outermost nodes in $A$ and $B$ followed by two nodes 1 lattice spacing closer to the boundary and so on till the nodes $1_A,2_A$ and $1_B,2_B$ are acted on. Finally there is a boundary interaction through $U_e$. In terms of the probability distribution according to which the edges are picked by the RQC the best sequence is a cycle of length(circuit depth) $k=L_A+L_B-1$ where the $(L_A-1)$ levels of the circuit choose edges in $A$ as follows: The first set of nodes $X_1$ is chosen as
\begin{align}
&\mathcal{P}^{(1)}(X_1)=\begin{cases}1~\text{iff}~|X_1|=2~\wedge~ D(X_1,1_A)=L_A-2\\=0~~~\text{otherwise}\end{cases}\nonumber\\
\end{align}
while the next $L_A-2$ levels choose edges depending on the choice of nodes in the previous step
\begin{align}
&\mathcal{P}^{(i)}(X_i|X_{i-1})=\begin{cases}1~\text{iff}~|X_{i-1}\cap X_i|=1~\wedge~\\~~~~~~~~~~~~~~~ D(X_i,1_A)=L_A-(i+1)\\=0~~~\text{otherwise}\end{cases}\nonumber\\
\end{align}
where $i=2,3,...,(L_A-1)$ and the distance between two sets $D(X_i,X_j)=min_x(|x_i-x_j|)$ is the minimum difference between any two elements belonging to the two different sets. In our case this difference is the difference of the position labels $(1,2,...,L_A)$. Similarly of the next $(L_B-1)$ levels of the circuit the first set of nodes $X_{L_A}$ in $B$ is picked as follows:
\begin{align}
&\mathcal{P}^{(L_A)}(X_{L_A}|X_{L_A-1})\nonumber\\
&~=\begin{cases}1~\text{iff}~|X_{L_A}|=2~\wedge~ D(X_{L_A},1_B)=L_B-2\\=0~~~\text{otherwise}\end{cases}\nonumber\\
\end{align}
while the next $(L_B-2)$ sets are chosen as 
\begin{align}
&\mathcal{P}^{(j)}(X_{j}|X_{j-1})\nonumber\\
&~=\begin{cases}1~\text{iff}~|X_{j}\cap X_{j-1}|=1~\wedge~ D(X_{j},1_B)=L_B-(j+1)\\=0~~~\text{otherwise}\end{cases}\nonumber\\
\end{align}
where $j=2,3,(L_B-1)$. Finally the RQC chooses the boundary edge as follows:
\begin{align}
&\mathcal{P}^{(L_A+L_B-1)}(X_{L_A+L_B-1}|X_{L_A+L_B-2})\nonumber\\
&~=\begin{cases}1~\text{iff}~|X_{L_A+L_B-1}|=2~\wedge~
|X_{L_A+L_B-1}\cap A,B|=1\\=0~~~\text{otherwise}\end{cases}\nonumber\\
\end{align}
One can similarly devise probability distributions for the RQC that generates any desired ordering of the edges.
However, for now, with the measure defined in Eq.(\ref{CEM}) and iterating the algebra of Eq.(\ref{singleedge}), we find that the purity is given by nested integrals:
\begin{align}
&\overline{P^{\alpha=2}}\!=\!\mathrm{Tr}[\omega^{\otimes 2}\!\!\int \!\! d[U_B]d[U_A]dU_e(U^{\dagger}_BU^{\dagger}_AU^{\dagger}_e)^{\otimes 2}T_A(U_eU_AU_B)^{\otimes 2}]\nonumber\\
&=\mathrm{Tr}[\omega^{\otimes 2}\int d[U_B]((U^{\dagger}_B)^{\otimes 2}\int d[U_A]((U^{\dagger}_A)^{\otimes 2}\nonumber\\
&~~~~~~~~~~~~~~~~~~~~~~~~~~~~~(\int dU_e (U^{\dagger}_e)^{\otimes 2}~T_A~U^{\otimes 2}_e)U^{\otimes 2}_A)U^{\otimes 2}_B)]\nonumber\\
&=\mathrm{Tr}[\omega^{\otimes 2}(N_d^2T_A+N_d^3T_{A-1}+N_d^4T_{A-2}..+N_d^{L_A}T_{A-L_A+2}\nonumber\\&~~~~~~~+N_d^{L_A}T_{A-L_A}+N_d^2T_A+N_d^3T_{A+1}+N_d^4T_{A+2}...\nonumber\\&~~~~~~~~~+N_d^{L_B}T_{A+L_B-2}+N_d^{L_B}T_{A+L_B})]\nonumber\\
&=(N_d^2+N_d^3+N_d^4+....+N_d^{L_A}+N_d^{L_A})+\nonumber\\&~~~~~~~~~~~~~~~~~~~~~~~~~~~~~~(N_d^2+N_d^3+N_d^4+....+N_d^{L_B}+N_d^{L_B})\nonumber\\
&=\frac{N_d^2(1-N_d^{L_A-1})}{(1-N_d)}+N_d^{(L_A-1)}\nonumber\\
&~~~~~~~~~~~~~~~~~~~~~~~~~~+\frac{N_d^2(1-N_d^{L_B-1})}{(1-N_d)}+N_d^{(L_B-1)}\nonumber\\
\label{bestcase}
\end{align}
The notation $T_{A+r}$ means the swap acting on $X=A\cup{1_B,2_B,...,r_B}$ and similarly $X=A-r=A\backslash{1_A,2_A,...,r_A}$. For $L_A,L_B$ reasonably large we find from eq.(\ref{bestcase}) that at the conclusion of the first cycle i.e. iteration 1, 
\be
\overline{P^{\alpha=2}}_{n_c=1}\approx \frac{2N^2_d}{(1-N_d)},
\ee
while for $n_c\ll L_A,L_B$
\be\label{bestcase2}
\overline{P^{\alpha=2}}_{n_c}\approx \left(\frac{2N^2_d}{(1-N_d)}\right)^{n_c},
\ee
and again we recall that this is for the best case sequence. Also
note that from Eq.(\ref{singleedge}) we see that for each non-trivial action of the averaging procedure over the unitaries we get a decrease of purity by the same amount,$(1-2N_d)$, thus a sequence which maximizes the number of non-trivial actions will result in the maximum decrease of purity. 

Now we want to compare the above case with the sequence that produces the least decrease in purity (or the least 2-Renyi entropy), or the {\em worst case}. We see that by keeping the same internal structure of $U_A,U_B$ and comparing the purities corresponding to the four possible cases: $(1)U=U_eU_AU_B,~(2)U=U_BU_eU_A,~(3)U=U_AU_eU_B,~(4)U=U_BU_AU_e$ we can easily see that for $n_c=1$ the sequence (4) is indeed the worst case scenario. Numerically we find that sequence (4) performs the worst also for generic large $n_c$. As for $n_c>1$, the decrease of purity also depends on what is the ordering of the unitaries in the products $U_A$ and $U_B$. We find numerically that the least decreasing sequence of unitaries (worst case) is given by choosing $U_{worst}=U_BU_AU_e$ with  $U_A=U_{a_{L_A-1}}....U_{a_2}U_{a_1}$ and $U_B=U_{b_{L_B-1}}....U_{b_2}U_{b_1}$. Let us quantify the purity using the worst sequence $U_{worst}$ as function of $n_c$. For $n_c=1$, we have 
\begin{align}
\overline{P^{\alpha=2}}_{n_c=1}&=\mathrm{Tr}[\omega^{\otimes 2}(N_dT_{A-1}+N_dT_{A-1})]=2N_d
\end{align}
We can also obtain the exact expression for the purity after any number of iterations $n_c\leq L_A$ where the size of the environment $L_B\geq L_A$. We find: 
\begin{align}
 &\bar{P^{\alpha=2}}_{n_c}\nonumber\\
&=\!\!\! \sum_{m=0}^{m=n_c-1}\!\!\mathrm{Tr}[\omega^{\otimes 2},\frac{C(n_c,m)}{2}N_d^{(n_c+m)}(T_{A-(m+1)}\!+\!T_{A+(m+1)})]\\
\label{progress}
&=2s^{n_c} \{1 - ((1 - N_d) N_d)^ {n_c}\binom{2 {n_c}-1}{ n_c}{}_2F_1(1, 2 n_c;1 + n_c; N)\}\nonumber\\
&={\frac{2N_d^{n_c}}{(1 - N_d)^{n_c}}} \{1 -f(n_c)\}
\end{align}
 where $C(n_c,m)=2\binom{n_c+m-1}{m},~ s=\frac{N_d}{(1-N_d)}$, and ${}_2F_1(1, 2 n_c;1 + n_c; N)$ is the Gauss Hypergeometric function. 
Therefore, our formalism allows us to obtain typicality of the purity for arbitrary depth of the RQC. When $n_c=O(1)$, the subsystem is very far from being the Haar case, and the RQC is not a $t-$design for any $t$, indeed, the system features an area law for the entanglement. As $n_c$ increases, entanglement propagates in the bulk at distance $\sim n_c$ from the boundary. For large values of $n_c$,  the expression simplifies because $f(n_c\to Large)\to 0$, and we see that 
 \be\label{worstpurity}
 \overline{P^{\alpha=2}}_{n_c\to Large}\approx \frac{2N_d^{n_c}}{(1-N_d)^{n_c}}.
 \ee
 Comparing this expression to the one for the purity in the best case sequence Eq.(\ref{bestcase2}), we see that, as long as $n_c$ is smaller than the system size, the best case sequence is better by a factor of $\frac{1}{2}(2N_d)^{n_c}$, see Fig.(\ref{asymptotic},top panel). Nevertheless, the same figure shows that after system size is reached, the two cases converge to a similar value. In the following, we show that the asymptotic value is independent of the ordering. This means that there is a $n_c$ above which there is an onset for the independence of the ordering. The numerical results shown in Fig.(\ref{asymptotic},top panel) suggest that the onset happens at $n_c\sim L_A$.

We can give a justification of why the above given sequences are indeed the ones that decrease the purity the most (best) or the least (worst). At $n_c=1$, of course, all the unitaries acting after $U_e$ are not entangling at all. With $n_c$ growing, all the unitaries that are acting \emph{before} $U_e$ allow some entanglement to be generated among the qubits. The sequences that go towards the boundary bring entanglement towards it, while the ones which start with the boundary and go outwards bring entanglement away. In this case, indeed the site $1_A$ would get very entangled with the bulk in $A$ (the same occurs to the $B$ side), and by monogamy this does not allow to effectively transfer entanglement information across the partition. This scenario also shows how in the worst case the RQC would entangle nodes at the same length as the iteration number.

At this point, we want to look for a result about the maximum purity of the $n_c-$iterated ensemble $\mathcal{E}^{n_c}$. The purity is a monotonically decreasing function because it is obtained by iterated application of CP maps, for all $n_c$.  As we pointed out above, the different results above hold in the region of validity for the scaling of $n_c$. Of course the absolute minimum of the purity of the reduced state cannot be less than $(\frac{1}{d})^{L_A}$ corresponding to the totally mixed state. To answer questions about the average distance of a state in $\mathcal{E}^{n_c}$ asymptotically to the totally mixed state on the subsytem $A$ we resort to the superoperator formulation a little later. As we noticed above, plot in Fig.(\ref{asymptotic},Top panel) also shows that for large $n_c$ the ordering does not count. This means that our scheme does indeed mimic the Trotter scheme as far as the statistics of the reduced system is concerned. In other words, the average decrease of purity in this model approximates the average decrease of purity obtained by evolving with local time dependent Hamiltonians and using the Trotter scheme. The irrelevance of ordering for large $n_c$ can be understood mathematically in the superoperator formalism, what we now go on describing for this model. 
\begin{figure}
 \centering
 \includegraphics[]{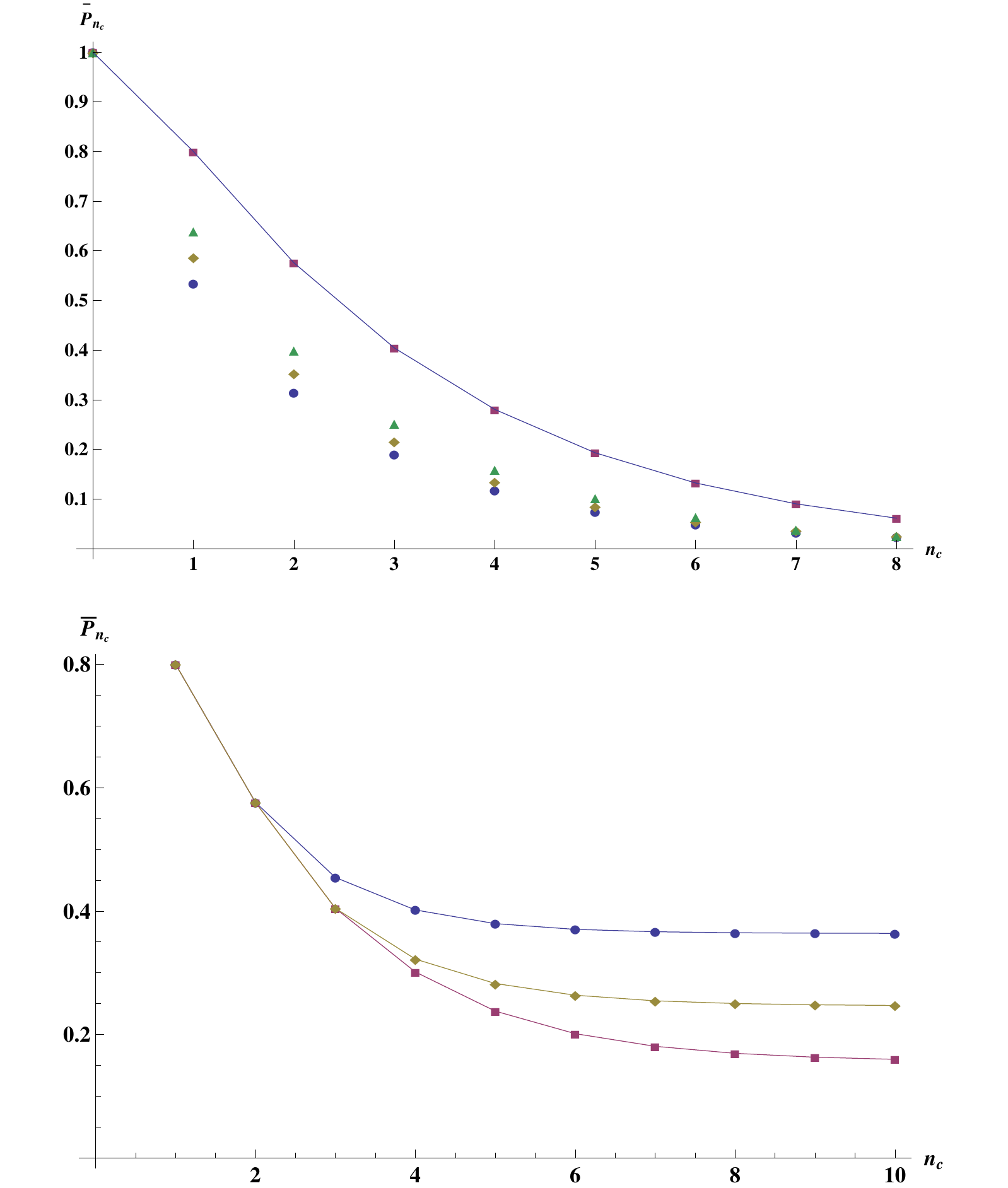}
 \caption{[\romannumeral 1](Top panel) Purity as a function of $n_c$ for the best sequence of unitaries(circles) and for the worst (squares) for $L_A=L_B=8$. The smooth line connecting the squares is the analytic expression for worst case $\overline{P}_{n_c}$. [\romannumeral 2] (Bottom panel) Convergence of the worst case sequence to the asymptotic formula value for different values of system/environment length. Blue circles($L=5,L_A=2$) and Yellow rhombi ($L=6,L_A=3$) converge to $>99\%$ of their asymptotic value whereas Purple squares($L=8,L_A=3$) converges to $>97\%.$}
  \label{asymptotic}
\end{figure} 
The chain superoperator is a $\sigma-$ordered product of non commuting projections i.e. $\mathcal{R}_{chain}=\mathcal{R}_{\sigma(|E|)}.....\mathcal{R}_{\sigma(1)}$, because of the non-commutavity of the local unitaries that make up the product. Infact the products in the superoperator is in reverse order of unitaries. For the sake of brevity without explicitly writing $\alpha=2$, using eq.(\ref{rsupop}) and the first line of eq.(\ref{bestcase}), one can see:
\begin{align}
U=U_{n}U_{n-1}....U_{1}\implies \mathcal{R}_{chain}=\mathcal{R}_{1}....\mathcal{R}_{n-1}\mathcal{R}_{n}
\label{rorder}
\end{align}
where the subscripts on $U$ denote supports for the same. Note that, for a fixed sequence, $\mathcal{R}_{chain}^{\dagger}\neq\mathcal{R}_{chain}$  and therefore  $\mathcal{R}_{chain}$ is not a hermitian operator. Nevertheless, the averaged sum over all possible sequences, $\mathcal{R}=\frac{1}{n!}\sum_{\sigma}\mathcal{R}_{\sigma}$, is Hermitian. Since our definition of the best sequence is $U=U_eU_AU_B$ with a specific \emph{internal} structure of unitaries within $U_A$ and $U_B$ and that for the worst sequence is $U=U_A^{\dagger}U_B^{\dagger}U_e$ Eq.(\ref{rorder}) implies that $\mathcal{R}_{best}=\mathcal{R}^{\dagger}_{worst}$. One can understand the action of the superoperator by studying it's action on the non-Orthnormal basis of the swap $T_X,~X\subset V$ operators. 
We regard the swap operators as kets in the $(|V|+1)-$dimensional subspace $\mathcal{S}=\mathrm{span}\{\ket{i}\},~0\leq i\leq |V|$ of the Hilbert-schmidt space on $(\mathcal{H}_V)^{\otimes 2}$. This subspace is the space of swap operators acting on all $i$ qubits from one end of the chain where $0\leq i\leq |V|$. The correspondence $\ket{i}=T_i=\openone_{V\backslash{i}}\otimes T_i$ then implies that the Hilbert-schmidt inner product $\braket{i}{j}\neq\delta_{i,j}$. In this basis the matrix representations of these superoperators are real and so it turns out that the eigenvalues of $\mathcal{R}_{best}$ and that of $\mathcal{R}_{worst}$ are identical. This means that for sufficiently large iteration number $n_c$ which sequence we choose does not matter while the rate of approach to the asymptotic value of purity is dictated by just the gap $(1-\lambda_2)$ in either case. The difference in initial decays of the purity for the sequences lies in the fact that the eigenvectors correspoding to identical eigenvalues are different.\\
\indent Let us now explain the action of $\mathcal{R}_{chain}$ in our chosen basis. Consider the action of the superoperator $\mathcal{R}_{chain}$ corresponding to the best sequence for the linear chain which has $|A|=L_A,~|B|=L_B$ when it takes as an argument some $T_X,~X\subset V$ which is the swap on $(\mathcal{H}_X)^{\otimes 2}$ subspace. Then the possibilities are :
\begin{align}
&\mathcal{R}_{chain}(\ket{0})=1,~~~~\mathcal{R}_{chain}(\ket{V})=1\nonumber\\
&\mathcal{R}_{chain}(\ket{i})=\sum_{p=0}^{i-1}N_d^{1+p}\ket{i+1-p}+N_d^i\ket{0},~~~1<i<L_A\nonumber\\
&\mathcal{R}_{chain}(\ket{L_A})=\sum_{p=1}^{L_A-1}N_d^{1+p}\ket{L_A+1-p}+N_d^{L_A}\ket{0}\nonumber\\
&~~~~~~~~~~~~~~~~~~+\sum_{p=1}^{|V|-L_A-1}N_d^{1+p}\ket{L_A-1+p}+N_d^{L_A}\ket{2L_A}\nonumber\\
&\mathcal{R}_{chain}(\ket{i})=\sum_{p=0}^{|V|-i-1}N_d^{1+p}\ket{i-1+p}+N_d^{|V|-i}\ket{V}\nonumber\\
&~~~~~~~~~~~~~~~~~~~~~~~~~~~~~~~~~~~~~~~~~~~~~~~~~, L_A<i<|V|\nonumber\\
\end{align}
In this basis the matrix representation $R$ of $\mathcal{R}_{chain}$ cannot be guaranteed to be even Normal thus we may only attempt a Jordan decomposition of $R$ but this is enough to understand the iterative behavior of the RQC. The purity of the generated ensemble after any number of iterations $k$ is given by:
\begin{align}
\overline{P^{\alpha=2}}_{n_c}&=\sum_{\ket{i}\in S}\bra{i}R^{n_c}\ket{L_A}
\end{align}
 From Eq.(\ref{rorder}) we see that $||\mathcal{R}_{chain}||\leq\prod_{e\in E}||\mathcal{R}_e||\leq 1$ which means that all eigenvalues $\lambda$, are less than equal to 1 in modulus and hence asymptotically only the contribution from fixed points survive. Note that each of the projections $\mathcal{R}_e$ has two eigenvectors $\openone$ and $T_V$ (which are swaps on no nodes and all nodes respectively), with eigenvalue 1. If one assumes then that the common eigenspace of $\mathcal{R}_{chain}$ is spanned by $\openone$ and $T_V$ we find that the asymptotic value of the purity is:
\begin{align}
\overline{P^{\alpha=2}}_{n_c\to\infty}&=\langle\omega^{\otimes 2},\mathcal{R}^{\infty}_{chain}(T_A)\rangle\nonumber\\
&=\langle\omega^{\otimes 2},\frac{\openone+T_V}{\sqrt{d^L(d^L+1)}}\rangle\langle\frac{\openone+T_V}{\sqrt{d^L(d^L+1)}},T_A\rangle+\nonumber\\
&~~~~~~~~~~~~~\overbrace{\langle\omega^{\otimes 2},\frac{\openone-T_V}{\sqrt{d^L(d^L+1)}}\rangle}^0\langle\frac{\openone-T_V}{\sqrt{d^L(d^L+1)}},T_A\rangle\nonumber\\
&=\frac{1}{\sqrt{d^L(d^L+1)}}\langle\frac{\openone+T_V}{\sqrt{d^L(d^L+1)}},T_A\rangle\nonumber\\
&=\frac{d^{2L-L_A}+d^{L+L_A}}{d^L(d^L+1)}
\label{asymp}
\end{align}
where the normalization for the symmetric and antisymmetric combinations of the basis vectors for the common eigenspace is obtained by setting $\langle\frac{\openone+T_V}{2C},\frac{\openone+T_V}{2C}\rangle=\mathrm{Tr}[\frac{\openone+T_V}{2C^2}]=1\implies C=\sqrt{d^L(d^L+1)}$.
We have verified this result numerically using small system lengths and the convergence with $n_c$ to the value given by eq.(\ref{asymp}) is shown by fig.(\ref{asymptotic}).

\begin{figure}
 \centering
 \includegraphics[width=\columnwidth]{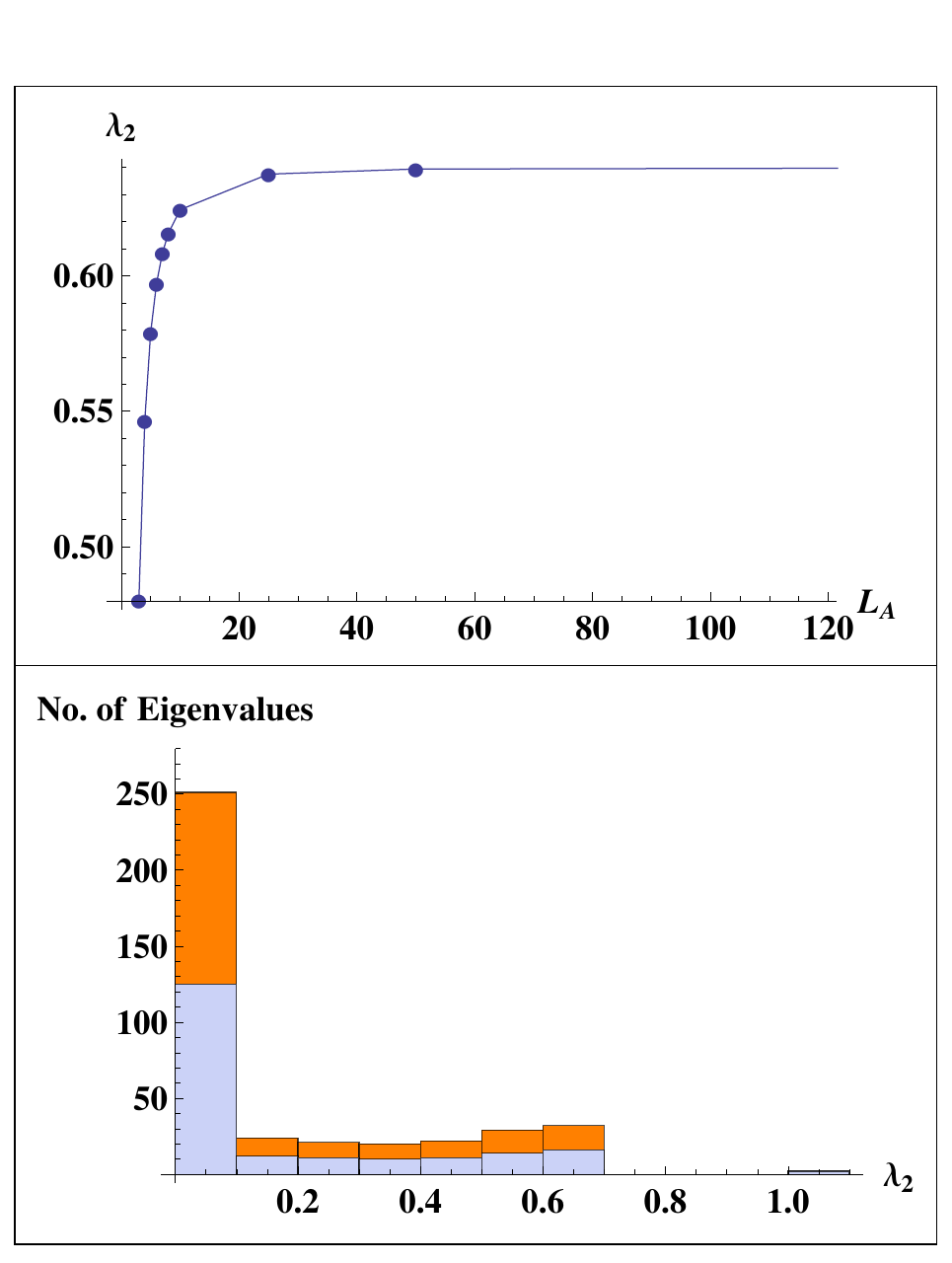}
 \caption{[\romannumeral 1]Saturation of the sub-dominant eigenvalue for $L_A=L_B$ with increasing $L_A$ for a chain of qubits. Similar behavior is observed for general qudits.[\romannumeral 2]Distribution of eigenvalues for a chain of qubits with $L_A=L_B$. The orange bars represent population levels within bands of the domains for the eigenvalues with $L_A=L_B=200$. Blue bars(superimposed on orange) are for $L_A=L_B=100$ The gap in the spectrum can be clearly seen as the difference between the largest eigenvalue of 1 and the next =.64}
  \label{lambda_saturation}
\end{figure}

For the specific case of $L_A=L_B$ we found numerically that the sub-dominant eigenvalue $\lambda_2$ saturates at  a value of $(2N_d)^2$ Fig.(\ref{lambda_saturation}) with increasing system size. Indeed numerical evidence indicates that the spectrum of $\mathcal{R}_{chain}$ at least for this case is entirely positive $0\leq {\lambda:\lambda\in\mathrm{Spec}(\mathcal{R}_{chain})}\leq 1$. All that happens upon increasing the system size is that the population of eigenvalues in the subregions of the domain $0\leq \lambda \leq \lambda_2$ increases proportionally, except for the largest eigenvalue of 1 whose number remains equal  to 2, see Fig.(\ref{lambda_saturation}). This corroborates our assumption that there exist only two fixed points. With this value of the asymptotic purity we can bound the average distance of states in our ensemble reduced to the subsystem, A, from the totally mixed state on it, $\sigma=\frac{\openone_A}{d^{|L_A|}}$. 

Indeed we see that the trace distance, $\overline{\mathcal{D}(\rho,\sigma)}=\frac{1}{2}\overline{||\rho-\sigma||_1}\leq \frac{1}{2}\sqrt{rank(\rho-\sigma)}\overline{||\rho-\sigma||_2}$ while
\begin{align}
&\frac{1}{2}\sqrt{\rho-\sigma}\overline{||\rho-\sigma||_2}=\frac{1}{2}\sqrt{\rho-\sigma}\overline{\sqrt{\mathrm{Tr}[\rho^2-\rho\sigma-\sigma\rho+\sigma^2]}}\nonumber\\
&~~=\frac{1}{2}\sqrt{d^{|A|}}\overline{\sqrt{\mathrm{Tr}(\rho^2)+\mathrm{Tr}(\sigma^2)-\frac{2}{d_A}}}\nonumber\\
&~~=\frac{1}{2}\sqrt{d^{|A|}}~~\overline{\sqrt{P(\rho)-P(\sigma)}}\leq\frac{1}{2}\sqrt{d^{|A|}}\sqrt{\overline{\overbrace{P(\rho)-P(\sigma)}^{\epsilon}}}\nonumber\\
&~~\leq \frac{1}{2}\sqrt{d^{|A|}}\sqrt{\bar{\epsilon}}\approx\frac{1}{2}\sqrt{\frac{d^{L_A}}{d^{L_B}}}
\end{align}
Thus for $d^{|B|}\gg d^{|A|}$ we can get indistinguishably close to the totally mixed state of the subsystem.

The analytical calculation of the variance for this model is difficult because the unitaries do not have disjoint supports. However, we can resort to a Markov type inequality for positive valued random variable to assert that the ensemble has a small variance since the average purity itself for large number of iterations reaches a value exponentially small in the system size.

\begin{figure}
 \centering
 \includegraphics[]{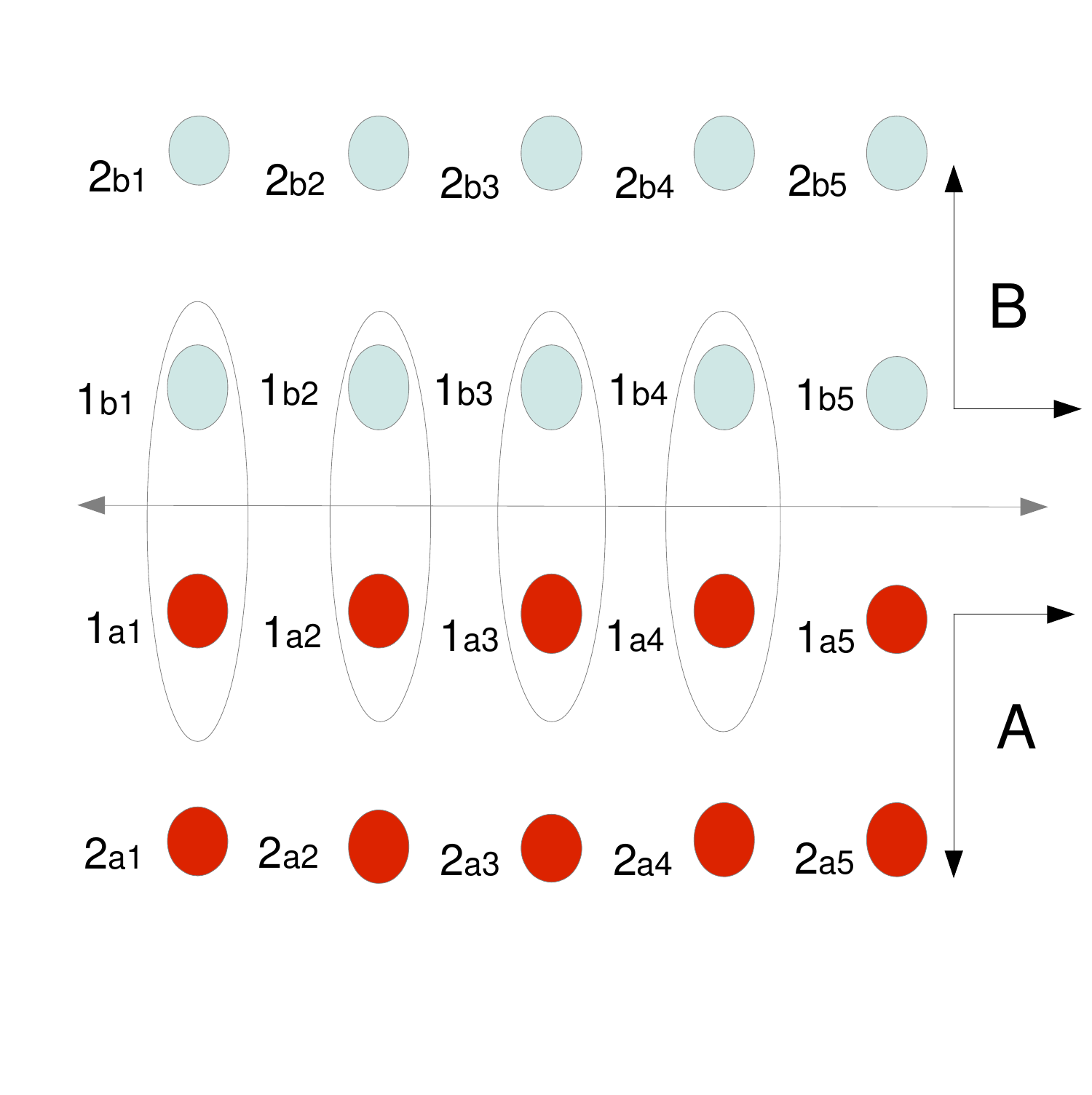}
 \caption{A bipartition $(A,B)$ of a square lattice.}
  \label{2d}
\end{figure} 

To summarize, the 1D model shows that for $n_c=O(1)$ almost every time dependent local Hamiltonian will evolve a factorized state to a state obeying the area law. The fluctuations around this average are small in view of the Markov inequality. On the other hand, when the number of iterations (the "time" in the Trotter scheme) scales with the size of the subsystem $n_c>L_A$, almost every evolution will produce a state with the volume law, and, asymptotically, maximally entangled. 

\subsection{Contiguous Edge Model on the Square Lattice}
Of course the area law in 1D is rather trivial. In order to find a more meaningful result, we need to study a two dimensional situation. We will see that in this case the fluctuations around the area law are even smaller.
We consider a 2-D square lattice with a bipartition into $A~\cup~B$ as in Fig.\ref{2d}. In this case the boundary between $A$ and $B$ is obviously a $1d$ system.  For sake of  simplicity,  let us start with  a RQC that applies Haar distributed 2-body gates across nodes of the bipartition that are nearest neighbors. It is easy then to see that  the unitaries across the boundary will have disjoint spaces of action and the average purity is a product of the single edge purities for the $n_c=1$ iterated ensemble. Explicitly ,
\begin{align}
&\overline{P^{\alpha=2}}=\!\mathrm{Tr}[\omega^{\otimes 2}\!\!\!\int dU_{e_1}(U^{\dagger}_{e_1})^{\otimes 2}T_{1a1}U_{e_1}^{\otimes 2}\!\!\int\!\! dU_{e_2}(U^{\dagger}_{e_2})^{\otimes 2}T_{1a2}U_{e_1}^{\otimes 2}\nonumber\\&~~~~~~~...\int dU_{e_l}(U^{\dagger}_{e_l})^{\otimes 2}T_{1al}U_{e_1}^{\otimes 2}T_{A\backslash{1a1,1a2,..,1al}}]\nonumber\\
&=\mathrm{Tr}[\omega^{\otimes 2}N_d(\mathbf{1}_{1a1,1b1}+T_{1a1}T_{1b1}).N_d(\mathbf{1}_{1a2,1b2}+T_{1a2}T_{1b2})\nonumber\\
&~~~~~~~~~~~~~....N_d(\mathbf{1}_{1al,1bl}+T_{1al}T_{1bl})T_{A\backslash{1a1,1a2,..,1al}}]\nonumber\\
&=\mathrm{Tr}[\omega^{\otimes 2}N^l_d(\mathbf{1}_{1a1,1b1,1a2,1b2,....1al,1bl}+\nonumber\\
&~~~~~~\mathbf{1}_{1a1}T_{1b1,1a2,1b2,....1al,1bl}+..\nonumber\\&~~~~~~~~~~~~+T_{1a1,1b1,1a2,1b2,....1al,1bl})T_{A\backslash{1a1,1a2,..,1al}}]\nonumber\\
&=(2N_d)^l
\end{align}
where $l$ is the number of closest nodes on either side of the boundary i.e. the length.This means that the 2-Renyi Entropy for this case can be lower bounded as $\overline{S}_2>-\mathrm{log}(\overline{P^{\alpha=2}})=-l~\mathrm{log}(2N_d)$. In words: the entropy is greater than a constant times the length of the boundary.In this case too it turns out that acting with the \emph{internal} unitaries first we can get lower purities and the reason is the same as the monogamy arguments presented for the linear chain analysis. For eg. in Fig.(\ref{2d}) if we consider the very simple system of just the nodes $1_{a1}~\text{and}~1_{a2}$ in $A$ and $1_{b1}~\text{and}~1_{b2}$ in $B$ then, upon implementation of the sequence $U=U_{1b1,1b2}U_{1a1,1a2}U_{1a1,1b1}U_{1a2,1b2}$ the average at iteration 1 is $\overline{P^{\alpha=2}}_{k=1}=(2N_d)^2=.64$ while for the sequence $U=U_{1a1,1b1}U_{1a2,1b2}U_{1b1,1b2}U_{1a1,1a2}$ it is $\overline{P^{\alpha=2}}_{k=1}=2N_d^2+8N_d^4=.525$.

Let us now compute the variance. For the 2-D model we find that since the edges have disjoint supports the average of the square of purity is a product of the average of squares of purity for each edge and similarly the average purity itself is a product of the average purity for each edge:
\ba
\overline{(P^{\alpha=2})^2_{2D}}&=\overline{(P^{\alpha=2})^2_{e_1}}~\overline{(P^{\alpha=2})^2_{e_2}}....\overline{(P^{\alpha=2})^2_{e_l}}\\
\overline{P^{\alpha=2}_{2D}}&=\overline{P^{\alpha=2}_{e_1}}~\overline{P^{\alpha=2}_{e_2}}....\overline{P^{\alpha=2}_{e_l}}
\ea
and the variance is thus
\begin{align}
Var[P^{\alpha=2}_{2D}]&=I^l-(2N_d)^{2l}\nonumber\\
&=\left(\frac{2(2d^4+9d^2+1)}{(d^2+3)(d^2+2)(d^2+1)}\right)^l-\left(\frac{2d}{(d^2+1)}\right)^{2l}
\end{align}
which, if we have qubits $d=2$ gives an exponentially decaying variance of
\be
Var[P^{\alpha=2}_{2D}]= - .64^l+.657^l
\ee
This is a strong result. The average 2-Renyi entropy does follow the area law. We are averaging over all the possible states obtained by local unitary transformations starting from a completely factorized state. This ensemble contains all the ground states of local Hamiltonian (without topological order). Moreover, we have shown that deviations from the area law are exponentially suppressed. Proving the area law in 2D for the ground states of local Hamiltonians is one of the most sought after results in quantum many body theory and our result represents a progress also in this direction.

\section{Conclusion and Discussion}
The question whether a physical model can generate an ensemble of states that reproduce moments over the Haar measure, that is, of the set of all the states in the Hilbert space, is 
an important question in quantum information theory, simulability of quantum many-body systems, and, recently the foundations of quantum statistical mechanics. In this paper we present some physical models with the aim of studying the statistical moments of the reduced subsystem $A$ of an initially bipartite quantum many-body system $A\cup B$, and with locality constraints that should contain the essence of the local dynamics induced by a local Hamiltonian. In this formulation, the relevant size $L_A$ at play is that of the subsystem $A$.

We present a superoperator theory that allows us to make statistical claims beyond the Haar measure. In particular, we are able to discuss both the typicality of area law (which is quite far from the Haar measure) and the volume law (which, indeed is what would hold for generic states in the Hilbert space). In simple words, we can show that states evolved with a local evolution for times of $O(1)$ with respect to $L_A$, exhibit \textit{typicality} of the area law for the entanglement, while for large times scaling with the subsystem size they exhibit a typical volume law. We want to stress the fact that typicality means that in the ensemble considered, the average is such and that the variance is vanishingly small in the large system limit. In other words, almost all members of the ensembles constructed show the above mentioned entanglement properties. In the asymptotic case, we recover results similar to those of $t-$designs, but again, where the scaling is that of the subsystem and not of the whole system. We have also discussed the implications of these findings for the foundations of statistical mechanics.

Our results are based on the algebra of the swap operators on subsets of the total nodes in the graph. This algebra shows how the underlying graph-theoretic structure of the system determines the propagation of entanglement within the subsystem. In particular, the mixing time for the subsystem is strongly dependent on the connectivity of the graph and other graph-theoretic notions like the average distance between two nodes or the Hausdorff dimension. Recently, it has been advanced that black holes are {\em fast scramblers}, that is, systems in which the mixing time is maximal \cite{ref43},\cite{ref44}, and several models for the fast scrambling process have been put forward \cite{ref43}-\cite{ref47}. We believe that our techniques can be useful also for this line of research. 

As we discussed in the introduction, as long as topological order is absent, ground states of local gapped Hamiltonians can be obtained by a circuit of fixed depth from a completely factorizable state. It would be interesting to study the statistics of the entanglement in ensembles where the fiducial state is topologically ordered. For instance, we would like to know if in such ensembles there is a non vanishing topological entropy \cite{ref48}-\cite{ref50} on average, and what are the fluctuations. This technique may prove useful to study the problem of the stability of topological phases.

\section{Acknowledgments}
PZ acknowledges partial support by the ARO MURI grant W911NF-11-1-0268 and NSF grants No. PHY-969969 and No. PHY-803304. This work was supported in part by the National Basic Research Program of China Grant 2011CBA00300 and 2011CBA00301 and the National Natural Science Foundation of China Grant 61073174, 61033001, and 61061130540. S.S. would like to thank the Perimeter institute for their hospitality during his visit where a large portion of this work was done. A special thanks to Dharmesh Jain, CNYITP, SUNY Stony Brook, for his help in writing and making some of the programs used for the numerics very efficient. Research at Perimeter Institute for Theoretical Physics
is supported in part by the Government of Canada through NSERC and by the Province of Ontario through MRI.

\appendix
\section{Calculation of product of traces usiing Schur's Orthogonality Theorem}
\label{A}
For the symmetric group  $S_{\alpha}$ of order $\alpha$, the elements are the permutation operators $\sigma$ on some space $\mathcal{H}$ where $\mathcal{H}$ can be broken up into a direct sum of Irreducible subspaces denoted by $\lambda$ each with dimension $d_{\lambda}$ and multiplicity $m_{\lambda}$ i.e. formally $\mathcal{H}=\oplus_{\lambda\in IRR(S_{\alpha})}\mathbf{C}^{m_{\lambda}}\otimes\mathbf{C}^{d_{\lambda}}$. For any operator $\overline{\sigma}$ on the same space we have that:
\begin{align}
\mathrm{Tr}(\sigma\overline{\sigma})&=\sum_{\lambda}m_{\lambda}\mathrm{Tr}_{d_\lambda}(\sigma\overline{\sigma})=\sum_{\lambda}m_{\lambda}\sum_{i,j=1}^{d_\lambda}\sigma^{(\lambda)}_{i,j}\overline{\sigma}^{(\lambda)}_{ji}\\
\mathrm{Tr}(\sigma)&=\sum_{\lambda'}\mathrm{Tr}_{d_{\lambda'}}(\sigma)=\sum_{\lambda'}\sum_{l=1}^{d_{\lambda}}\sigma^{(\lambda')}_{ll}
\end{align}

Using the above 2 equations we get that:
\begin{align}
&\frac{1}{\alpha!}\sum_{\sigma}\mathrm{Tr}(\sigma\overline{\sigma})\mathrm{Tr}(\sigma)\nonumber\\
&=\frac{1}{\alpha!}\sum_{\sigma\in S_{\alpha}}\sum_{\lambda,\lambda'\in IRR(S_{\alpha})}m_{\lambda}m_{\lambda'}\times\sum_{i,j=1}^{d_{\lambda}}\sum_{l=1}^{d_{\lambda'}}\sigma_{i,j}^{(\lambda)}\sigma_{ll}^{(\lambda')}\overline{\sigma}^{(\lambda)}_{ji}\nonumber\\
&=\sum_{\lambda,\lambda'}m_{\lambda}m_{\lambda'}\sum_{i,j,l}\overbrace{(\frac{1}{\alpha!}\sum_{\sigma\in S_{\alpha}}\sigma^{(\lambda)}_{i,j}\sigma^{\lambda'}_{ll})}^{\frac{1}{d_{\lambda}}\delta_{\lambda,\lambda'}\delta_{i,l}\delta_{j,l}}\overline{\sigma}^{(\lambda)}_{ji}\nonumber\\
&=\sum_{\lambda}\frac{m^2_{\lambda}}{d_{\lambda}}\sum_{i=1}^{d_{\lambda}}\overline{\sigma}^{(\lambda)}_{ii}=\sum_{\lambda}\frac{m^2_{\lambda}}{d_{\lambda}}\chi^{(\lambda)}(\overline{\sigma})
\end{align}
where in the second line above the Kronecker deltas appear due to Schur's great Orthogonality theorem \cite{ref40}.

\section{Calculation of Variance}
\label{B}
The trace w.r.t. $\Pi_+$ in Eq.(\ref{ranvariance}) can be given a closed form expression again by using Schur's Orthogonality relation for representative functions on the irreps of a group see for eg.\cite{ref40}). The result is 
\be
\mathrm{Tr}[\Pi_+(T^{(12)}_i\otimes T^{(34)}_i)=\sum_{\lambda:\text{IRRs of} S_4 }\frac{m^2_\lambda}{d_\lambda}\mathrm{Tr}[(T^{(12)}_i\otimes T^{(34)}_i)\Pi_\lambda].
\ee
Notice that now the sum is over just the irreps of $S_4$. The symmetric group of order 4, $S_4$ has $4!=24$ elements which are all permutation operators $\sigma$ on 4 labels. In our case the labels refer to copies of the $i$'th or $j$'th space that make up $(\mathcal{H}_i\otimes\mathcal{H}_j)^{\otimes 4}$ . Thus $\sigma=\sigma_i\otimes\sigma_j$ is a valid decomposition of the permutation operators. The projector onto the totally symmetric part of $(\mathcal{H}_i\otimes\mathcal{H}_j)^{\otimes 4}$ takes the form $\Pi_+=\frac{1}{24}\sum_{\sigma\in S_4}\sigma_i\otimes\sigma_j$ and hence
\begin{align}
& \mathrm{Tr}_{i,j}[(T^{(12)}_i\otimes T^{(34)}_i)\Pi_+]\nonumber\\
&~~~~~~~~~~~~~~~~~~~~~=\frac{1}{24}\sum_{\sigma\in S_4}\mathrm{Tr}_{i,j}[(T^{(12)}_i\otimes T^{(34)}_i\otimes\openone_j)(\sigma_i\otimes\sigma_j)]\nonumber\\
&~~~~~~~~~~~~~~~~~~~~~=\frac{1}{24}\sum_{\sigma\in S_4}\mathrm{Tr}_{i}[(T^{(12)}_i\otimes T^{(34)}_i.\sigma_i)].\mathrm{Tr}_{j}[(\openone_j.\sigma_j)]
\end{align}
Note that in cycle notation the operator $T^{(12)}_i\otimes T^{(34)}_i=(12)_i(34)_i$. Thus it's product with any other $\sigma_i$ still gives us a permutation operator on the $i$ space. One can then just count the number of cycles in the product obtained where each cycle contributes a multiplicative factor of the dimension $d$ of the space. For e.g $\mathrm{Tr}_i(\sigma_i=(12)_i(34)_i)=d^2,~\mathrm{Tr}_i(\sigma_i=\openone_i)=d^4$ since there are two cycles in the permutation $(12)_i(34)_i$ and 4 in $(\openone_i)$. Similarly the traces over the $j$th spaces can also be obtained.
Finally the calculation above yields $\mathrm{Tr}_{i,j}[(T^{(12)}_i\otimes T^{(34)}_i)\Pi_+]=\frac{d^2(2d^4+9d^2+1)}{12}$


\end{document}